\documentclass[aps,twocolumn,prd,superscriptaddress,10pt,nofootinbib]{revtex4-1}

\usepackage{graphicx,amssymb,amsmath,amsthm,amsfonts,epsfig,epsf,fixmath}
\usepackage{bm}
\usepackage{dcolumn}
\usepackage{latexsym}
\usepackage{rotating}
\usepackage{longtable}
\usepackage{enumerate}
\usepackage{tensor,multirow}
\usepackage{xurl}
\usepackage{comment}
\usepackage[linktocpage]{hyperref}
\usepackage[usenames]{color}
\usepackage{epstopdf}
\usepackage{tikz,tikz-3dplot}
\usepackage{aas_macros}
\usepackage{ulem}
\usepackage{slashed}
\usepackage[scr=boondox]{mathalfa}
\usepackage[scr=boondox]{mathalfa}
\usepackage{slashed}
\usepackage{ulem}
\usepackage{tikz,tikz-3dplot}
\usetikzlibrary{positioning,arrows.meta,calc}
\usepackage[bottom]{footmisc}

\DeclareFontFamily{U}{min}{}
\DeclareFontShape{U}{min}{m}{n}{<-> udmj30}{}

\begin{document}
\title{Gravitational waves from parabolic encounters: \\ A study of linear and non-linear memory}

\author{Samik Dutta}
\email{samik.dutta@edu.ufes.br}
\affiliation{PPGCosmo, Universidade Federal do Espírito Santo, 29075-910, Vitória, ES, Brazil}
\affiliation{Birla Institute of Technology and Science - Pilani, Rajasthan, 333031, India}

\author{Ankur Chhabra}
\email{ankurchhabra.physics@gmail.com}
\affiliation{Birla Institute of Technology and Science - Pilani, Rajasthan, 333031, India}

\author{Aritra Banerjee}
\email{aritra.banerjee@pilani.bits-pilani.ac.in}
\affiliation{Birla Institute of Technology and Science - Pilani, Rajasthan, 333031, India}
\affiliation{Asia Pacific Center for Theoretical Physics, Postech, Pohang 37673, Korea}
 
\author{Sajal Mukherjee}
\email{sajal.mukherjee@pilani.bits-pilani.ac.in}
\affiliation{Birla Institute of Technology and Science - Pilani, Rajasthan, 333031, India}

\author{Subhendra Mohanty}
\email{subhendram@iiserb.ac.in}
\affiliation{Department of Physics, IISER Bhopal, Madhya Pradesh, 462066, India}

\date{\today}

\begin{abstract}
The memory effect is known to introduce a permanent displacement in the gravitational wave (GW) detectors after the passage of a GW signal. While the \textit{linear memory} adheres to the source properties, the \textit{non-linear memory} is a secondary effect sourced by the GW itself. In the present work, we discuss GW signals with both these kinds of memory effects, while focusing on the parabolic limit of an encounter. This special case is theoretically intriguing and emerges as a limiting situation for both eccentric and hyperbolic events. However, in this paper, we argue that a simple extrapolation of memory calculations for eccentric or hyperbolic cases to the parabolic case may lead to incorrect estimations. Therefore, we treat the parabola as a special case and use an intrinsic parameterization, with which we calculate gravitational wave signals and their energy spectrum via an effective field theory formalism. Unlike the hyperbolic case, which is known to have linear memory, we notice that parabolic encounters bring out new features in the zero frequency limit (ZFL). The exactly parabolic case is studied here primarily as an idealized separatrix between bound and unbound motion, and our analysis highlights some of the key challenges and salient aspects of GW memory in this regime. 
\end{abstract}

\maketitle
\flushbottom

\definecolor{napiergreen}{rgb}{0.16, 0.5, 0.0} 

\renewcommand*{\l}{\lambda_{\star}}
\newcommand*{\M}{M_{\star}}

\section{Introduction \& Motivation \label{sec:introduction}}

The detection of GWs by the LIGO-Virgo-Kagra (LVK) collaboration \cite{LIGOScientific:2016aoc} has opened a new window to explore the \textit{universe}. It provides an excellent scope for testing predictions of Einsteinian gravity \cite{LIGOScientific:2021sio, PhysRevD.103.122002, universe7120497}, and possibly constrain alternatives to \textit{general relativity} (GR) \cite{PhysRevD.108.024027, Arun13:altgr}. While it is of interest to continue exploring events such as mergers with present detectors \cite{Abac_2025}, there is significant motivation to model relativistic effects that are prime targets for future GW detectors \cite{PhysRevD.107.064048, PhysRevD.110.122006}. GW \textit{memory} \cite{Zeldovich:1974gvh, Braginsky:1985vlg} is one such possible effect that could serve as a potential source for upcoming detectors like Laser Interferometer Space Antenna (LISA) \cite{LISA:2022kgy, Inchauspe:2025, Ghosh_2023}, Cosmic Explorer (CE)\cite{LIGOScientific:2016wof, Grant:2022bla}, and Einstein Telescope (ET)\cite{Hild:2010id, Goncharov:2023woe}. A GW signal with nonzero memory manifests as a permanent relative displacement of two detectors at early and late times \cite{Favata_2010}. Such a signal is characterized by the difference between asymptotic values of the strain amplitudes in particular polarizations: 
\begin{equation}
  \label{deltah}
    \Delta h_{+, \times}^{mem} = \lim_{t \to \infty} h_{+, \times}(t) - \lim_{t \to -\infty} h_{+, \times}(t), 
\end{equation}
and a nonzero value of the above would render the fact that there is a permanent separation before and after the wave passes, and hence a nonzero memory!
\par
The GW memory is primarily categorized into two, namely, linear and non-linear. While linear memory is present in specific GW sources, in contrast, the non-linear memory can be associated with any GW source. For example, a binary on an unbound/hyperbolic orbit, with sources that go through non-oscillatory motion, can emit GWs with nonzero linear memory. Similarly, asymmetry can also introduce a GW memory, which can often be found in \textit{supernova} explosions \cite{Burrows:1995bb, Kotake:2005zn, Murphy:2009dx}, neutrino emission bursts \cite{Epstein:1978dv, Turner:1978jj}, kicks \cite{Favata:2008ti} and gamma-ray burst jets \cite{Takashi:2005, Sago:2004, Segalis:2001}. Interestingly, these effects pertaining to linear memory will affect the emitted GW, and may cause a dephasing \cite{10.1093/mnras/stz2925, Vartanyan_2020}. On the other hand, non-linear memory (or Christodoulou memory) \cite{Blanchet:1992br, Payne, Christodoulou:1991cr}  is a second-order effect and originates due to the stress-energy tensor of GWs themselves, more precisely due to the self-interaction thereof \cite{Misner:1973prb}. The stress energy tensor is proportional to the GW energy radiation, and therefore, the perturbation too is proportional to GW radiation. It turns out that the oscillatory part of the non-linear GW appears at a 2.5 PN order, while the non-oscillatory part kicks in at an order higher than that \cite{Favata:2011qi}. There has been some recent progress in this particular aspect, particularly in detecting memory effects through GW observations \cite{Inchauspe:2025, PhysRevD.101.083026, Ghosh_2023}. The upcoming detectors, such as LISA, CE, and ET, are expected to achieve sensitivities that could discern non-linear memory from binary mergers \cite{Goncharov:2023woe, Grant:2022bla}. Recent works \cite{Inchauspe:2025, Cogez:2026frh} show the numerical and statistical analyses of GW memory detection with LISA. If detected, distinctly from other non-linear effects, non-linear memory will definitely shed light on processes involving graviton-graviton coupling or the sheer non-linear nature of Einstein's field equations\footnote{For such linear/non-linear memory waveforms used in the context of detectors, one can see \cite{Detect1, Detect2, Detect3}. For numerical relativity (NR) based approaches, the reader is referred to \cite{Pollney:2010hs, Mitman:2020bjf} and references therein.}.
\par
With the above discussion stored as background, we can now realize that unbound sources like hyperbolic encounters do have both linear and non-linear memory \cite{Favata:2011qi, Hait:2022ukn}. However, the bound eccentric orbits, on the other hand, only have non-linear memory and no linear memory component \cite{Favata:2011qi, Hait:2022ukn}, a fact easily attributed to the symmetry in velocity asymptotes. In this paper, we discuss a very specific case, i.e., what happens when the eccentricity becomes $1$, and we have parabolic encounters! Several studies expand on the formation of binaries through captures \cite{PhysRevD.102.083016, Gondan:2020svr, Gond_n_2022}, while marking the importance of the parabolic limit \cite{Bae:2017crk, Gond_n_2018}. These events may provide a closer look at the strong field regime since they represent marginally bound orbits where one compact object may momentarily pass very close to another compact body. However, the formation of a binary system at eccentricity exactly equal to 1 is a fine-tuned physical system. The motion of a realistic astrophysical binary is more likely to be bound \cite{Peters:1964, PetersMathews:1963}, or unbound/hyperbolic \cite{Mukherjee:2020hnm}, rather than exactly parabolic \cite{Berry:2010gt, Bae:2017crk}. 
\par
In the present context, our motivation is primarily theory-driven and two-fold. First, we would like to have a deeper look at the linear memory component in parabolic encounters. To be more precise, we would first like to understand how eccentricity ($e$) ranges from hyperbolic signals with memory ($e>1$) to the eccentric ones without memory ($e<1$), and especially how orbits behave at the $e \rightarrow 1$ limit. It turns out that the last bit has some very subtle difficulties to be achieved smoothly. In doing so, one would also like to take a look at an approach to memory grounded in Quantum Field Theory (QFT), viz. the \textit{soft-graviton theorem \cite{weinberg1965photons, weinberg1965infrared} which connects low frequency graviton emission phenomena to the memory effect}. Specifically, the zero frequency regime of the amplitude has a pole in frequency space sitting inside the kinematic factor that relates the hard and soft amplitudes, and this manifests as an irreversible step in the Fourier space \cite{strominger2016gravitational}. However, direct computations from this perspective also have some subtlety since it assumes the presence of well separated asymptotic states. We thus pursue calculating memory directly via a frequency space derivation of the energy radiated, by using the stress tensors of the parabolic orbits as a source and effectively computing the tree level amplitude. The other/second motivation of the paper is to work out the non-linear memory contribution to the parabolic encounters, which one does by using emitted gravitons themselves as a source.  One would expect non-linear memory to be manifested in the trajectory, possibly close to the periapsis, where the interaction is strongest. This may also trigger events like capture\footnote{Note that other types of orbits (in the probe particle limit), including a zoom-whirl one \cite{Martel:2003jj} could be possible in this case. But we will not consider these very special scenarios.}, resulting in an eccentric and bounded orbit, which will be the focus of our work. 
\par
The rest of the paper is organized as follows: In section-(\ref{sec:framework}), we introduce the motivation and framework to study parabolic encounters with relevant parameterization of the orbit and a perspective on the ZFL for this case. Next, section-\eqref{sec:effectiveft} is devoted to introducing the preliminary effective field theory aspects of the frequency space computation. In section-(\ref{sec:linearmem}), we discuss the linear memory for this binary, which we find to be zero as expected. We then proceed to introduce the radiated power estimation in such binaries in section-(\ref{sec:radiation}), since they have a burst-like structure and may lead to dynamical capture scenarios. In section-(\ref{sec:nonlinmem}), we discuss the non-linear memory and extensively focus on qualitative features of the non-linear signal. Finally, in section-(\ref{section6}), we end the paper with comments and future directions. Some mathematical details of the calculations are provided in the appendices. Throughout this paper, we work in geometrized units, i.e., $G = c = 1$. Also, all physical quantities are made dimensionless by normalising them with the total mass $M$, except in section-(\ref{sec:radiation}) where we compare our results to \cite{Berry:2010gt}.
\section{Subtleties of the parabolic encounter \label{sec:framework}}
\subsection{Discontinuity from limits}
To begin our discussion, recall geometrically that parabolic encounters can be thought of as a limiting case of both elliptical and hyperbolic orbits. Naturally, it may be tempting to believe that results for parabolic events can be obtained as a limiting case of $e \lessgtr 1$. In this section of the paper, we argue whether this idea is viable, particularly in terms of memory calculations. From the seminal works of Zeldovich \cite{Zeldovich:1974gvh}, Thorne \cite{PhysRevD.45.520, 1987Natur.327..123B}, and others \cite{Braginsky:1985vlg, Christodoulou:1991cr, Blanchet:1992br, PhysRevD.44.R2945, Favata:2011qi,bieri2014perturbative}, we know that hyperbolic orbits can contain nonzero linear memory. In a naive sense, the linear memory is associated with the incoming and outgoing velocity asymptotes. When these asymptotes have a non-vanishing difference, we obtain a nonzero value of linear memory. It turns out that this limit is closely related to the zero frequency limit (ZFL) \cite{Smarr}. This is the reason that a hyperbolic encounter can have nonzero radiation even when the frequency is approaching zero.
\par
Let us now revisit the hyperbolic encounter and directly apply the $e \rightarrow 1^+$ limit there. By following Refs. \cite{1977ApJ...216..610T, Landau:1975}, a hyperbolic orbit confined on $x-y$ plane can be parametrized as follows
\begin{equation}
\begin{alignedat}{2}
    x(\xi)&=a(e - \cosh\xi),   &y(\xi)&=b \sinh \xi,\\
    z(\xi)&=0,  & \omega^{\prime} t/\nu &=\omega_0 t=(e\sinh \xi - \xi).
\end{alignedat}
\label{eq2}
\end{equation}
where $a$ and $b$ are semi-major and semi-minor respectively, $\omega_0$ is the natural frequency. If we now focus on the relation between $\omega^{\prime}$ and $\omega_0$, we notice that $\nu=\omega^{\prime}/\omega_0$ is a dimensionless quantity that we want to put to zero\footnote{Note that for a given hyperbolic orbit of fixed $e$, the ZFL corresponds to taking the reference frequency $\omega' \to 0$.}. \newline The natural frequency can be written as $\omega_0=G^{1/2}M^{1/2}a^{-3/2}$, where $M=m_1+m_2$ is the total mass of the binary. By using the relation between $a$ and $b$, that is $b = a\sqrt{e^2-1}$, the frequency $\omega_0$ becomes $\omega_0 = G^{1/2}M^{1/2}b^{-3/2}(e^2-1)^{-3/2}$. Therefore, as we are approaching the parabolic limit, we seemingly have $\nu \rightarrow \infty$. Now this potentially leads to a problem to extend the hyperbolic limit to a parabola, because we are interested in finding the $\nu \rightarrow 0$ limit or ZFL whenever memory effects are concerned. This problem could be better elucidated from the expressions of the stress-energy components for binaries in a hyperbolic encounter. We have discussed it in greater detail in section \ref{sec:linearmem}.
\par
It is also possible to approach the parabola limit from an eccentric orbit. The parameterization for eccentric orbits reads as \cite{Landau:1975, Berry:2010gt}
\begin{equation}
\begin{alignedat}{2}
    x(\xi)&=a(\cos\xi-e), \quad&y(\xi)&=b \sin\xi,\\
    z(\xi)&=0,\quad &\omega_{n}t/n&=\omega_0 t=(\xi-e \sin\xi).
\end{alignedat}
\end{equation}
In the above expression, $\omega_n$ is the frequency corresponding to the $n$-th harmonic. Here also, we notice that the parabolic limit $e\to1^-$ will be given as $n \rightarrow \infty$, i.e., harmonic expansions valid for the elliptic case fail to converge as we take the parabolic limit. Therefore, the basic framework from both sides breaks down as we approach a parabolic encounter. This can, however, be understood as a problem with the coordinate system we are working with since it renders time to periastron effectively infinite. 

\subsection{A new parameterization \label{subsection 2.1}}
\begin{figure*}[t]
\centering
\begin{tikzpicture}[scale=0.9,>=stealth]
%
    \draw[<->] (-3.5,0) -- (1.5,0) node[right] {$x$};
    \draw[<->] (-1,-2.0) -- (-1,2.0) node[above] {$y$};

    \draw[thick, domain=-1.7:1.7, smooth, variable=\y] 
        plot ({-(\y)^2}, \y);
        
    \fill[black] (-1,0) circle (0.05) node[above left] {$m_1$};
    \fill[black] (-0.25,-0.5) circle (0.04) node[below right] {$m_2$};
    
    \filldraw[black] (-1,0) circle (0.03) node[below left] {Focus};

    \filldraw[black] (0,0) circle (0.01) node[below right] {Vertex};

    \draw[dashed, gray] (-3.5,-1.87) -- (0,0);
    \draw[dashed, gray] (-3.5, 1.87) -- (0,0);

    \node[gray, left] at (-3.5,1.87) {\textbf{Asymptote}};
    \node[gray, left] at (-3.5,-1.87) {\textbf{Asymptote}};

    \node at (-5.0,0.3) {Semi-major axis ($a=\infty$)};
    \begin{scope}[xshift=2.8cm, yshift=2.0cm, scale=1.4]
        \draw[thick, gray] (-0.6,-2.0) rectangle (2.2,2.0);

        \draw[<->] (-0.5,0) -- (1.8,0) node[right] {$x$};
        \draw[<->] (0,-1.5) -- (0,1.5) node[above] {$y$};

        \fill[black] (0,0) circle (0.05) node[below left] {$m_1$};

        \def\rp{1.0} 
        \def\numax{83}
        \draw[->, thick,domain=-\numax:\numax,smooth,variable=\phi,samples=200]
          plot ({(2*\rp)/(1+cos(\phi)) * cos(\phi)},
                {(2*\rp)/(1+cos(\phi)) * sin(\phi)});

        \fill[black] (\rp,0) circle (0.04);

        \node[below right=1pt] at (\rp,0) {\scriptsize Periapsis};

        \node[below right=8pt] at (\rp,0) {\scriptsize $(\xi=0)$};

        \draw[<->] (0,0.15) -- (\rp,0.15) node[midway,above] {$r_{\min}$};

        \def\nuex{-60}
        \coordinate (P) at ({(2*\rp)/(1+cos(\nuex)) * cos(\nuex)},
                            {(2*\rp)/(1+cos(\nuex)) * sin(\nuex)});
        \fill[black] (P) circle (0.04) node[below right] {$m_2$};

        \draw[dashed] (0,0) -- (P) node[midway,below left] {$r$};

        \draw[->, thick] (0.5,0) arc[start angle=0,end angle=\nuex,radius=0.5];
        \node at (0.65,-0.25) {$\phi$};

        \node[black] at (1.5,-1.75) {$\xi=\tan\frac{\phi}{2}$};
    \end{scope}
\end{tikzpicture}
\caption{Schematic of a parabolic trajectory for a binary system involving masses $m_1$ and $m_2$. Inset: we focus on the region near the periapsis, which will be our main interest.}
\label{traj}
\end{figure*}
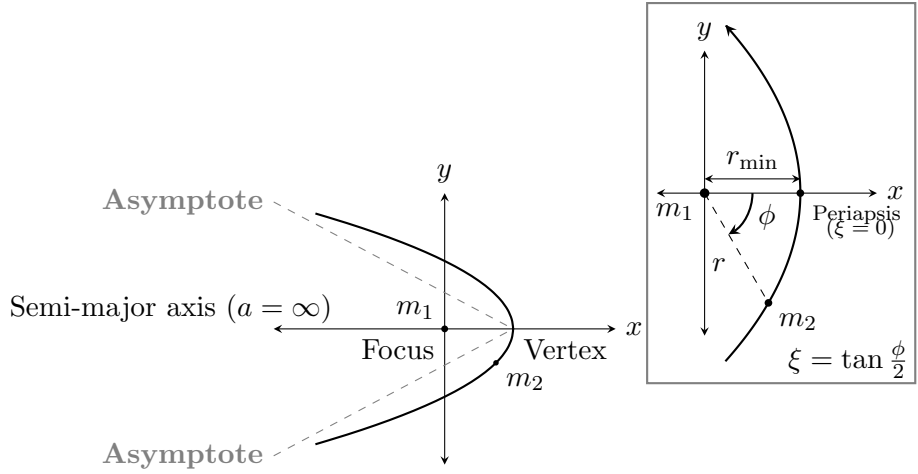

With the discussions above, we now have an intuition as to why the orbital parameterizations for Keplerian bound orbits $(e < 1)$ or for unbound orbits $(e > 1)$ are unfit to describe a parabolic encounter of binary black holes. Therefore, we employ a separate  parameterization for the specific case when $e = 1$. Such a parameterization was introduced in Ref. \cite{1977ApJ...216..610T}, following which we consider a black hole of mass $m_2$ in a parabolic trajectory around another black hole of mass $m_1$, as shown in Fig. \eqref{traj}. Working in the center-of-mass frame and reducing the system to a single body with reduced mass $\mu = \frac{m_1 m_2}{M}$, the orbit is defined in terms of the mean eccentric anomaly $\xi$ as follows\footnote{For the parabolic case, the period is infinite, so one can't define the mean eccentric anomaly the usual way, but an analogous effective description can be given. Note that the near periapsis region is given by $|\xi|\to 0$ (or $\phi \to 0$). From scaling of time with $\xi$, one could infer that motion is almost linear in $\xi$ near the periapsis region, but away from it, the $\xi^3$ term dominates, signifying a slow approach phase.}:
\begin{equation}
\label{eq:parameterization}
     \begin{split}
       &\xi = \tan{\left(\frac{\phi}{2}\right)},  
       \quad r = r_{\rm min}(1 + \xi^2), 
       \quad t = 2 \tau \left(\frac{\xi^3}{3} + \xi\right),  \\
    &\cos\phi = \frac{1-\xi^2}{1+\xi^2}, 
       \quad \sin\phi = \frac{2\xi}{1+\xi^2},
\end{split}
\end{equation}
where $\phi$ is the true anomaly, i.e. the actual angular position of the reduced mass system when measured from the periapsis of the parabola, $r_{\rm min}$ is the distance from periapsis, and $\tau$ is the characteristic time scale of the encounter. Hence, the coordinate parameterization for the orbital trajectory in Cartesian system in 2D becomes dependent on $\xi$ as,
\begin{align}
    &\vec{l}(\xi) = (x(\xi), y(\xi),z(\xi)) \nonumber \\
    & = (r\cos\phi, r\sin\phi, 0) = (r_{\rm min}(1-\xi^2), 2r_{\rm min}\xi, 0),
\end{align}
and the velocity vector becomes,
\begin{equation}\label{velvec}
    \vec{u}(\xi) = \frac{d\vec{l}}{dt} = \frac{d\vec{l}}{d\xi}\frac{d\xi}{dt}=\left(\frac{-\xi r_{\rm min}}{\tau(1+\xi^2)}, \frac{r_{\rm min}}{\tau(1+\xi^2)}, 0 \right). 
\end{equation}
It is easy to see from the magnitude of this velocity that it decreases with $1/|\xi|$ (or with $r^{-1/2}$) away from the periapsis position. The fundamental frequency and the angular frequency of the orbit are related as in the case before: $\omega' = \nu\omega_{0}$, where $\nu \in$ (0, $\infty$) is a non-negative dimensionless real number and, 
\begin{equation}\label{tauscale}
    \omega_{0} = \frac{1}{\tau} = \left(\frac{2GM}{r^3_{min}}\right)^\frac{1}{2},
\end{equation}
has been evaluated at $r_{\rm min}$ since that is the characteristic length scale for the timescales of the encounter\footnote{For example, in bound Keplerian orbits the semi-major axis $a$ plays the role of the length scale, but that is not usable in the parabolic case. The periapsis distance plays the role of the finite length scale in this situation.}. 

\subsection{The soft theorem perspective}

In the case of unbound binary encounters, the soft-graviton theorem provides a powerful way to compute memory effects via scattering amplitude techniques. The core idea of using soft graviton theorems \cite{weinberg1965photons, weinberg1965infrared} to get GW waveforms can be suggestively written as the scaling of the perturbed metric in terms of the retarded time at a detector\footnote{See Ref. \cite{Sen:2024qzb} for an excellent recent review.}. In frequency space, the same can be written as:
\begin{equation}
    h_{\mu\nu} \sim \frac{A_{\mu\nu}}{\omega}+ B_{\mu\nu}\ln\omega +...
\end{equation}
where the coefficients $A, B$, etc. can be calculated from the knowledge of incoming and outgoing momentum states only, and not on the details of the process itself. Here, clearly the $\mathcal{O}(\frac{1}{\omega})$ term gives one the leading memory term in the limit $\omega\to 0$, and the pole can be clearly calculated from the leading soft factor associated with the radiated soft graviton in an $n-$particle scattering process \cite{weinberg1965infrared}:
\begin{equation}
   \frac{4G}{r} \left( \sum_{\lambda=1}^{n} \epsilon^{\lambda}_{\alpha\beta}\epsilon^{*\lambda}_{\mu\nu} \right)\left(\sum_{a=1}^{n} \frac{\eta_{a}p^{\mu}_{a}p^{\nu}_{a}}{p_{a}.k} \right)
\end{equation}
where, we have a soft graviton of momenta $k^{\mu}$, polarization $\epsilon_{\alpha\beta}^{\lambda}$ and the hard particles interacting with each other have momenta $p_a^{\mu}$. For incoming particles  the symbol $\eta_{a} = -1$ and for outgoing particles $\eta_{a} = +1$. Note that we only want to focus on the $\frac{1}{\omega}$ pole coming from the external leg emission and are not including any subleading terms involving angular momentum. For the parabolic case, without going into the simplification of the soft factor, one can readily see just by using the parameterization in Eq. \eqref{eq:parameterization} and Eq. \eqref{velvec} that the spatial in/out velocities on the $x-y$ plane
\begin{align}
u_{x}^{in}(\xi\rightarrow -\infty)&=0; \qquad u_{x}^{out}(\xi\rightarrow \infty)=0,\\
u_{y}^{in}(\xi\rightarrow -\infty)&=0; \qquad u_{y}^{out}(\xi\rightarrow \infty)=0.
\end{align}
We can see that the asymptotic velocities for the parabolic trajectories are zero, and hence all components of $A$ are zero:
\begin{equation}
    A_{xx}=A_{yy} = A_{xy} = 0.
\end{equation}
So clearly, there should be no linear memory term in this case. One could similarly calculate $B$, the loop-level $\mathcal{O}(\ln \omega)$ tail term from the subleading soft theorem \cite{Sen:2024qzb, Laddha:2018myi, Sahoo:2018lxl, Hait:2022ukn}, and in the parabolic case, it also identically turns out to be zero due to the vanishing asymptotic velocities.
\par
However, herein lies the subtlety in using the universal soft theorem for the special case of a parabola. Since the particle starts and ends at rest at the asymptotes, there is no long-range momentum transfer to contribute to a soft graviton pole, contrary to a hyperbolic case. The caveat is that for a marginally bound trajectory, the assumption of well defined asymptotic states breaks down, as all states degenerate to zero momentum at infinity. This implies that parabolic motion may be outside of the strictly defined universal sectors of the soft theorem regime, and there may be some non-universal terms appearing in this case. We will see it later in section-(\ref{sec:linearmem}), that $\omega \rightarrow 0$ limit in the parabolic case brings out a new pole structure, which was not reported earlier in the literature from soft theorem perspective.
\par
With this, we end the discussion on the subtleties and possible signatures of the parabolic encounters in the soft theorem perspective. In the upcoming sections, we carry out the frequency space calculation for linear and non-linear memory in the parabolic case, and whenever required, we will compare with known/expected results.

\section{Effective field theory and memory effect \label{sec:effectiveft}} 
In an effective field theory approach to gravity, the Effective Field Equations (EFE) are derived by considering fluctuations of the background metric as a graviton field $h_{\mu \nu}$ \cite{Mohanty:2022abo}. We begin with the Einstein-Hilbert action:
\begin{equation}
    \mathcal{S} = \int d^4 x \sqrt{-g} \left[-\frac{1}{16 \pi G}\right] R, 
\end{equation}
where, $g = \det(g_{\mu \nu})$, $R$ is the Ricci scalar and $G$ is Newton's gravitational constant. In the weak field approximation, the metric tensor $g_{\mu \nu}$ is expanded around the flat Minkowski spacetime $\eta_{\mu \nu}$ as
\begin{equation}
    g_{\mu \nu} = \eta_{\mu \nu} + \kappa h_{\mu \nu},
\end{equation} 
where $\kappa = \sqrt{32 \pi G}$. We would now like to see how GWs can be expressed via the stress tensor of the source written in momentum space. The probability amplitude of a graviton emitted by a source with a stress-energy tensor $\Tilde{T}^{\mu\nu}(k)$ in momentum space (defined via Fourier transform) is written as \cite{Mohanty:2022abo}
\begin{equation}
    \label{prob amplitude}
    \mathcal{A}_{\lambda}(k_{0}, \vec{n}k_{0}) = -i\frac{\kappa}{2}\epsilon^{*\lambda}_{\mu\nu}(\vec{n})\Tilde{T}^{\mu\nu}(k_{0}, \vec{n}k_{0}).
\end{equation}
Here, $\vec{n}$ is a directional unit vector and $\epsilon^{*\lambda}_{\mu\nu}$ are polarization tensors for the spin-2 field. The gravitational perturbation metric can be written in terms of this probability amplitude as
\begin{equation}
\begin{split}
    \label{h in terms of A}
        h_{\alpha\beta}(\vec{x}, t) = \frac{1}{4 \pi r}\int\frac{dk_{0}}{2\pi}\sum_{\lambda=1}^{2} & \epsilon^{\lambda}_{\alpha\beta}(\vec{n})\mathcal{A}_{\lambda}(k_{0}, \vec{n}k_{0})\\ 
        & \exp\left(-i k_{0}(t-r)\right),
\end{split}
\end{equation}   
and therefore, the perturbation metric can be identified as GW as $\Tilde{h}_{\mu\nu} \equiv g_{\mu\nu} - \eta_{\mu\nu} = \kappa h_{\mu\nu} $, giving the following relation
\begin{equation}
\begin{split}
\label{tildeh in terms of A}
        \Tilde{h}_{\alpha\beta}(\vec{x}, t) = \frac{\kappa}{4 \pi r}\int\frac{dk_{0}}{2\pi}\sum_{\lambda=1}^{2} & \epsilon^{\lambda}_{\alpha\beta}(\vec{n})\mathcal{A}_{\lambda}(k_{0}, \vec{n}k_{0})\\ & \exp\left(-i k_{0}(t-r)\right).
\end{split}
\end{equation}
Here, the graviton field $\Tilde{h}_{\alpha \beta}$ is a canonical spin-2 field with mass dimension 1, and $r$ is the distance to the source.  Substituting Eq. \eqref{prob amplitude} into Eq. \eqref{tildeh in terms of A}, gives the waveform at the detector
\begin{eqnarray}
    \tilde{h}_{\alpha\beta}(\vec{x}, t) = -\frac{4G}{r}\int\frac{dk_{0}}{2\pi}\Big( \tilde{T}_{\alpha\beta}(k_{0}, \vec{n}k_{0}) -\nonumber \\
    \frac{1}{2}\eta_{\alpha\beta} \tilde{T}^\mu_\mu(k_{0}, \vec{n}k_{0}) \Big)\exp\left(-i k_{0}(t-r) \right),
\end{eqnarray}
where the completeness relation 
\begin{equation}
\sum_{\lambda=1}^{2}\epsilon^\lambda_{\mu\nu}(k)\epsilon^{*\lambda}_{\alpha\beta}(k) = \frac{1}{2}(\eta_{\mu\alpha} \eta_{\nu\beta} + \eta_{\mu\beta} \eta_{\nu\alpha}) - \frac{1}{2}\eta_{\mu\nu}\eta_{\alpha\beta},    
\end{equation}
is employed. Upon using the projection operator $\Lambda_{ij,kl}(\vec{n})$ to obtain the Transverse-Traceless (TT) component of the GW metric, we obtain the gauge-fixed expression in terms of the spatial component:
\begin{align}
    [\tilde{h}_{ij}]^{TT}(\vec{x},t) = -\frac{4G}{r} \Lambda_{ij,kl}(\vec{n}) \int \frac{dk_{0}}{2\pi} T_{kl} (k_{0}, \vec{n}k_{0})\nonumber\\
    \exp\left( -i k_{0}(t-r)\right),
\end{align}
which in the frequency space can be written as:
\begin{equation}
\label{h in freq space}
     [\Tilde{h}_{ij}]^{TT}(\vec{x},k_{0}) = -\frac{4G}{r}\Lambda_{ij,kl}(\vec{n}) T_{kl}(k_{0}, \vec{n}k_{0}),
\end{equation}
or the amplitude for each polarization can equivalently be written in terms of the contraction of the gauge-fixed stress tensor with the polarization basis: 
\begin{equation}
\label{eq:hlambda}
    h_{\lambda}(\omega', r) = \frac{4G}{r} \epsilon^{ij}_{\lambda} (\vec{n}) T_{ij} (\vec{n}, \omega'),
\end{equation}
where $\lambda$ is the polarization of the signal and $\hat{n}$ is the radial direction of the observer.
Considering the graviton as a quantum field and deriving the radiated energy  in terms of source stress-energy tensors, the expression reads:
\begin{align}
\label{energy radiated field theory}
    E_{gw} = \frac{\kappa^2}{4} \sum_n \int \frac{8 \pi}{5} \left(T_{ij}(\omega_n') T^*_{ji}(\omega_n') - \frac{1}{3}|T^i_{ i}(\omega_n')|^2 \right) \nonumber \\
    \omega^3 2 \pi \delta(\omega_n' - \omega) \frac{d\omega}{(2 \pi)^3 2 \omega}.
\end{align}
This expression is for a general source with a stress tensor $T_{ij}(\omega_n')$ in Fourier space, and can be used to compute the energy radiated by a binary. The memory effect can be obtained from soft-graviton amplitudes by considering an $n$-body scattering, whereas the GW from scattering amplitude is given by \cite{Mohanty:2022abo}:
\begin{align}
    \label{gw from scattering}
    \Tilde{h}^{TT}_{ij}(\vec{n}r, t) = \frac{4G}{r} \int \frac{dk_{0}}{2\pi i k_{0}} \sum_{a=1}^{n} \frac{m_{a}}{\sqrt{1-v_{a}^2}}\left[\frac{v_{ai}v_{aj}}{(1-\vec{v}_{a}\cdot\vec{n})}\right]^{TT} \nonumber \\
    \exp\left(-i k_{0}(t-r) \right),
\end{align}
where $m_{a}$ and $v_{a}$ are the mass and asymptotic velocity of the (hard) particles respectively, the expression in the brackets is actually nothing but the soft factor, and $\hat{n}$ gives the direction of the detector from the source. 
Since the integral can be written in terms of the Heaviside function $\Theta(t-r)$, the change in separation of a detector of two masses after $t > r$, when the GW with memory passes through it, is given by a DC offset induced by the soft factor:
    \begin{align}
       & \Delta h^{TT}_{ij} = h_{ij}(t >> r_{0}) - h_{ij}(t << r_{0}) \nonumber \\
       & = \frac{4G}{r} \sum_{a=1}^{n}\frac{m_{a}}{\sqrt{1-v_{a}^2}}\left[ \frac{v_{ai}v_{aj}}{(1-\vec{v}_a\cdot\vec{n})}\right]^{TT},
    \end{align}
where $r_{0}$ is the separation between the two masses of the detector. This separation of the test masses is irreversible. To see this clearly, recall that the manifestation of the linear memory effect in the GW signal takes place as a pole in the $\Tilde{h}_{ij}(\omega)$ if the signal in the Fourier space is of the form $\Tilde{h}_{ij}(\omega) = {A}/{\omega}$  \cite{Hait:2022ukn, Mohanty:2022abo}. In time, this corresponds to a memory waveform of the form:
\begin{equation}
    h(t) = \int^{\infty}_{-\infty} \frac{d\omega}{2 \pi} e^{i \omega t} \frac{A}{\omega} = A \Theta(t),
\end{equation}
where $\Theta(t)$ is the Heaviside step function. 
For instance, for hyperbolic orbits, where particles at past and future infinities could have very different velocities, the energy radiated at zero angular velocity is non-zero. This is a signature of the GW memory effect in the zero-frequency dominated waveform, which, as we will witness in the next section, vanishes in the marginally bound limit $e \rightarrow$ 1.
\par
The \textit{non-linear} or Christodoulou memory, on the other hand, results from the radiative mass multipole moments which are sourced by the energy-flux of the radiated primary GWs, hence making it a secondary effect \cite{Blanchet:1992br, Christodoulou:1991cr}. Consider the relaxed EFE $\square\Tilde{h}^{\alpha\beta} = - 16 \pi \tau^{\alpha\beta}$, where the effective stress-energy tensor $\tau^{\alpha\beta}$ comprises of the matter stress-energy tensor $T^{\alpha\beta}$, Landau-Lifshitz (LL) pseudotensor $t^{\alpha\beta}_{LL}$, and terms quadratic in $\Tilde{h}^{\alpha\beta}$. This LL pseudotensor of the primary source contains non-linear terms of the order $(\partial h)^2$ and serves as the source for non-linear GWs satisfying a reduced EFE  \cite{Favata_2010}. Following Refs. \cite{PhysRevD.45.520, PhysRevD.44.R2945}, one finds that the non-linear part, corresponding to secondary radiation, arises as the correction to the GW field (by solving for the Green's function of reduced EFE):
\begin{equation}
    \label{nonlinear correction}
    \delta h^{TT}_{jk} = \frac{4G}{r} \int_{-\infty}^{T_{r}}  \left[\int \frac{dE^{gw}}{dt'd\Omega'}\frac{n'_{j}n'_{k}}{(1-\vec{n}'\cdot\vec{n})}\, d\Omega'\right]^{TT}\, dt',
\end{equation}
where $T_r = t - r$ is the retarded time, $r$ is the distance of the observer from the source, $\hat{n}$ is the unit vector along the direction of emission of the primary graviton and $\hat{n}'$ being the unit vector along the direction of emission of the secondary graviton. The time integral in the above equation gives the memory component its hereditary nature, i.e. the quantity depends on the entire burst history of the radiative source and its dynamics. In the TT gauge, the non-linear memory waveform finally reads \cite{Hait:2022ukn, Mohanty:2022abo}:
\begin{equation}
\label{eq:nonlinear wavefunction}
    \left[h_{ij}^{mem} \right]^{TT} = \frac{4G}{r} \int^{T_r}_{-\infty} dt' \int_{4\pi} d\Omega' \frac{dE_{gw}}{dt' d\Omega'} \times \frac{\Lambda_{ij, kl}(\hat{n}') n_k' n_l'}{1 - \hat{n}' \cdot \hat{n}},
\end{equation}
where again, $\Lambda_{ij, kl}$ is the transverse-traceless projection operator\footnote{Note that the GW energy flux can usually be written in terms of the square of the Bondi news.}. This will be the key equation which we will use in the subsequent sections. Note that in the above expression, the non-linear memory signal originates from the primary gravitational waves through the stress tensor interaction term, which reads $T_{ij}T_{kl}^* \Lambda_{ij, kl}$.

\section{Linear GW memory from parabolic orbits \label{sec:linearmem}}
Let us start by applying the formalism discussed in the last section and calculating the linear memory (soft scattering) term for the parabolic encounter. We have already mentioned that for the marginally bound case, the linear memory should just vanish via symmetry considerations; this is what we will be elucidating on next. 
\par
Considering a system of compact binaries as point masses, the action for such a system is given by \cite{Landau:1975, Mohanty:2022abo}, 
\begin{equation}
    S = -\sum_n m_n \int d\tau_n = -\sum_n m_n \int \sqrt{g_{\mu \nu} l_n^{\nu} l_n^{\mu}}.
\end{equation}
The stress energy tensor for these system of masses $m_n$ moving along a worldline $z^{\mu}(\tau)$ is, 
\begin{equation}
    T^{\alpha \beta}(z^{\mu}) = \sum_n \int  m_n u_n^{\alpha} u_n^{\beta} \delta^4(l^{\mu} - z_n^{\mu}(\tau_n)) d\tau_n,
\end{equation}
where the 4-velocity is given as $u^{\alpha} = \dfrac{dl^{\alpha}}{d\tau}$. When we parameterise the worldline in terms of the coordinate time $t$ so that $l^i = l^i(t), \vec{u} = \dfrac{d\vec{l}}{dt}$, then
\begin{equation}
    T^{\alpha \beta}(t, \vec{l}) = \sum_n m_n \frac{u_n^{\alpha} u_n^{\beta}}{u_n^t} \delta^3 (\vec{l} - \vec{z}_n(t)).
\end{equation}
In the center of mass frame of the binary, with the reduced mass $\mu$, the Fourier transform of the stress energy tensor in frequency space can be written as \cite{Mohanty:2022abo}:
\begin{equation}
\begin{split}
        T_{xx}(\omega') &= \int_{-\infty}^{\infty} e^{i\omega't} \mu \dot{x}^2 dt, \\
        &= -i\mu\omega'\int_{-\infty}^{\infty} dt e^{i\omega't}x\frac{dx}{dt},\\   
        &= -i\mu\omega'\int_{-\infty}^{\infty} x d\xi e^{2i\nu(\frac{\xi^3}{3}+\xi)}\frac{dx}{d\xi}.\\
\end{split}
\end{equation}
Note that the second expression is obtained by applying integration by parts and ignoring a term proportional to $\Ddot{x}(t)$. Finally, the $xx$-component of the stress-energy tensor (per unit reduced mass $\mu$) in terms of $\xi$ is written as:
\begin{equation}
\label{eq:Txx-unsolved}
    T_{xx}(\omega') = 2i \omega' r_{\text{min}}^2 \int^{\infty}_{-\infty} \xi (1 - \xi^2) e^{2i \nu \left(\frac{\xi^3}{3} + \xi \right)} d\xi.
\end{equation}
Similarly, the remaining stress-energy components are:
\begin{align}
\label{eq:Tyy-unsolved}
    T_{yy}(\omega') = -4i \omega' r_{\text{min}}^2 \int^{\infty}_{-\infty} \xi e^{2i \nu \left(\frac{\xi^3}{3} + \xi \right)} d\xi,\nonumber \\
    T_{xy}(\omega') = 4i \omega' r_{\rm min}^2\int^{\infty}_{-\infty} \xi^2 e^{2i \nu \left(\frac{\xi^3}{3} + \xi \right)} d\xi.
\end{align}
The time parameterization in the parabolic case makes sure we have the $\xi^3$ term in the frequency integral, that dominates near the periapsis, since that is where most of the radiating event takes place. It is well-known that the integrals of these forms are computable in terms of Airy functions and their derivatives\footnote{See Appendix \ref{sec:Airy} for a discussion on Airy functions and their application to solve integrals of this form.}. Once the integrals are solved, the final expressions for the stress-energy components, in terms of $\nu$, take the form:
\begin{align}
\label{eq:Txyfinal}
    T_{xy}(\nu) &= -2^{8/3} \pi i \frac{\nu^{2/3}}{\tau} r_{\text{min}}^2 \left[Ai((2 \nu)^{2/3}) \right],\\
\label{eq:Tyyfinal}
  T_{yy}(\nu) &= -\pi 2^{7/3} \frac{\nu^{1/3}}{\tau} r_{\text{min}}^2 \left[Ai'((2 \nu)^{2/3})\right],\\
\label{eq:Txxfinal}
    T_{xx}(\nu) &= \frac{\pi r_{\text{min}}^2}{\tau}  \Big[ 2^{7/3} \nu^{1/3} Ai'((2 \nu)^{2/3}) + \nonumber \\
     & 2^{2/3} \nu^{-1/3} Ai((2 \nu)^{2/3}) \Big],
\end{align}
where we have substituted $\omega^{\prime}=\nu \omega_{0}$, and Eq. (\ref{tauscale}). Now remember that a signature of the memory effect is the contribution of the waveform dominating at zero frequency where $\nu \rightarrow 0$. This is manifested in the form of a $\frac{1}{\nu}$ pole in frequency space, which gives rise to a Heaviside function in the time domain. This leads to a permanent, non-oscillatory displacement in the position of test particles following the passage of gravitational waves.
\par
Before we proceed with the $\nu \rightarrow 0$ limit for the parabolic case, let us quickly revisit the same for a hyperbolic event. To do this, we use the same parameterization as in Eq. \eqref{eq2} and follow Ref. \cite{Hait:2022ukn}. The reason to study this case is to approach $e \rightarrow 1$ limit from the hyperbolic side, and observe how the energy-momentum tensors behave in the parabolic limit. For a binary system undergoing a hyperbolic encounter, one can work out the source stress-energy tensors (see Eq. (30) of \cite{Hait:2022ukn}) as:
\begin{equation}
\begin{aligned}
    T_{xx} &= -\frac{2 \mu a^2 \omega_0}{e^2}\left[\ln{(\nu e)} + (e^2 - 1) \right], \\
    T_{yy} &= -\frac{2\mu a^2 \omega_0}{e^2} (e^2 - 1) \left[\ln{(\nu e)} - 1 \right], \\
    T_{xy} &= 2i\mu \nu \omega_0 a^2 \sqrt{e^2 - 1} \left[\frac{e^2 - 1}{e^2} \ln{(\nu e)} - \frac{1}{\nu^2 e^2} \right]. \\
\end{aligned}
\end{equation}
Here, if we take the $e \rightarrow 1$ limit then $T_{xy} = T_{yy} = 0$. The only term that survives is
\begin{equation}
    T_{xx} = -2\mu a^2 \omega_0 \ln (\nu).
\end{equation}
Now if we approach the $\nu \rightarrow 0$ limit, this term diverges. Therefore, there is a clear discontinuity when you try to approach the parabolic limit ($e \rightarrow 1$) and ZFL ($\nu \rightarrow 0$) simultaneously from the hyperbolic side. This strongly calls for the new parameterization in Eq. \eqref{eq:parameterization} to study the special case of parabolic encounters.
\par
\begin{figure*}[htp]
    \centering
    \includegraphics[width=0.65\textwidth]{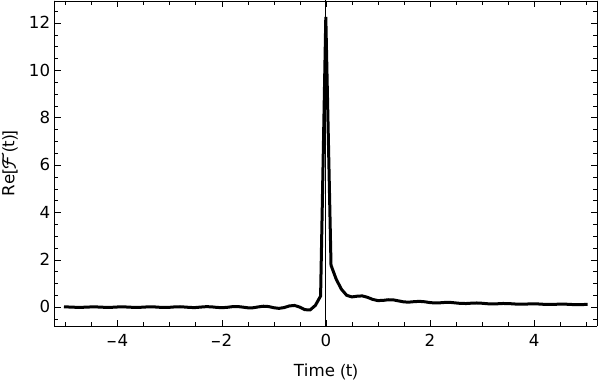}
    \caption{Behaviour of $\displaystyle \frac{1}{\nu^{1/3}}$ pole in the time domain, with the transient spike at $t=0$, obtained through numerical Fourier transform.}
    \label{fig:ParabolicFT}
\end{figure*}

With this detour, let us now come back to the parabolic encounter and deal with equations \eqref{eq:Txyfinal}-\eqref{eq:Txxfinal}. Turns out, when we expand the Airy functions and their derivatives (see Appendix \ref{sec:Airy}) around $\nu \to 0$, no $\frac{1}{\nu}$ poles are present in the expressions of the stress energy tensors $T_{xy}$, $T_{yy}$  and $T_{xx}$. Although the leading order expansion of Eq. \eqref{eq:Txxfinal} at the ZFL gives,
 \begin{equation}\label{fracpole}
     \lim_{\nu \to 0} T_{xx}(\nu) \simeq\frac{\pi r^2_{min}}{\tau} \frac{2^{2/3}}{3^{2/3} \Gamma(2/3)} \frac{1}{\nu^{1/3}} + \cdot \cdot \cdot
 \end{equation}
 This very intriguing pole $\frac{1}{\nu^{1/3}}$ in the second term of the component $T_{xx}$ is nothing but a remnant of the Airy function asymptotics\footnote{A similar scaling can be identified in \cite{Berry:2010gt} as well by carefully looking at the Airy-Bessel relations.} This term does not contribute to linear memory from binaries in parabolic orbits because it does not produce a Heaviside function in the time domain. There is a spike seen around $t = 0$ and then the growth and decay on either side goes as $t^{-2/3}$, as shown in Fig. \eqref{fig:ParabolicFT}, until there is no permanent distortion (when $t \rightarrow \pm \infty$) as opposed to what is seen with a Heaviside function. Note that this kind of fractional pole is absent from the universal structure of soft graviton theorem, and can only be seen as a classical stationary phase contribution. The presence of this pole, a particular signature for this case, reaffirms our earlier discussion on the uniqueness of ZFL for parabolic binaries. See Appendix \ref{sec:appB} for a detailed discussion of this pole from a systematic limit of the hyperbolic case. 
\par
In conclusion, there is no linear memory resulting from binaries in a parabolic encounter, confirming the already expected result. However, radiation is emitted over a finite time near the periapsis, and afterwards the system eventually returns to the pre-encounter state asymptotically, i.e., no permanent displacement.
\par
From an observation point of view, the slowly decaying term in the leading order is a distinctive feature, and may serve as an explicit indicator of the parabolic encounter as it impacts the low-frequency domain of the wave. In contrast, for hyperbolic encounters, the leading late-time behavior is dominated by a constant term, i.e., the memory effect, while any decaying tail arises from the subleading contributions. Of course, a potential caveat is whether this tail lies within the sensitivity of the GW detectors or not. In practice, much of the asymptotic tail is expected to fall below the low-frequency threshold of the present ground-based detectors. However, the early-time portion of the tail, immediately following the burst, may have some potential relevance within the scope of present GW detectors. Nonetheless, the future low-frequency observations, especially from LISA and pulsar timing arrays (PTAs), may be more sensitive to this regime, where both the tail and memory effects have greater observational relevance \cite{Inchauspe:2025, Cao:2026, Cogez:2026frh, Zosso:2026}. A more robust estimation of the detectability of these events is left for future work.

\section{Radiation during parabolic encounter \label{sec:radiation}}
From what we have seen so far, parabolic encounters are intrinsically different from eccentric binary evolution, and it is evident from their radiation properties. Unlike binaries with $e<1$, parabolic/hyperbolic encounters have a burst-like radiation, which is spanned over a very small amount of time, and in addition, centered exactly around the periapsis. In this section, we will discuss these aspects and also try to understand how intrinsic binary parameters can be affected due to radiation. Of course, these parameters will later be connected with non-linear memory in the next section.
\subsection{Computation of energy radiation}
For the computation of the energy radiation, we will first consider the quadrupole formula, and then use the field theoretic calculations. 
\subsubsection{Using quadrupole formula:}
\begin{figure*}[htp]
  \centering  \includegraphics[width=0.65\textwidth]{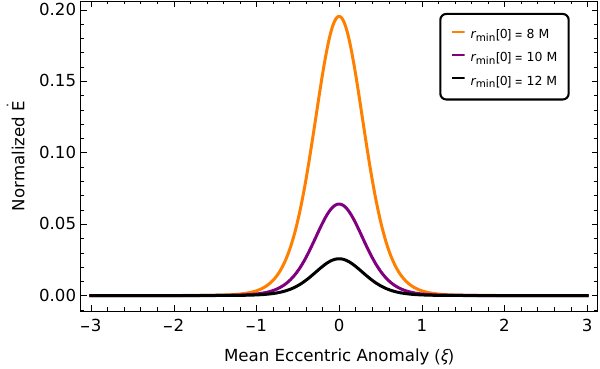} 
  \caption{Power Spectrum for different $r_{\rm min}$ values.}
  \label{fig:Power spectrum - variable Lz}
\end{figure*}

Let us now obtain the power radiation from the quadrupole formula given as \cite{maggiore2008gravitational, Landau:1975}
%
\begin{equation}
    \label{eq:power equation}
    P = -\frac{dE}{dt} = \frac{1}{5} \left(\dddot{Q}^2_{11} + \dddot{Q}^2_{22} + 2\dddot{Q}^2_{12} + \dddot{Q}^2_{33} \right), 
\end{equation}
where $Q_{ij}$ represents the quadrupole moment written as $Q_{ij}=\mu \left(x_i x_j -r^2\delta_{ij} /3\right)$, with $\delta_{ij}$ being the Kronecker delta function. Using the parameterization in Eq. \eqref{eq:parameterization} and following the prescription laid out in Refs. \cite{Peters:1964, PetersMathews:1963}, the power radiated in a parabolic encounter is calculated to be:
\begin{equation}
\label{eq:power}
    P = \frac{16 M^3 \mu^2 }{15 r^{5}_{min} (1 + \xi^2)^{6}} (12 + \xi^2),
\end{equation}
in terms of the mean eccentric anomaly $\xi$. In Fig. \eqref{fig:Power spectrum - variable Lz}, we have plotted the power for different $r_{\rm min}$, reaffirming that the power is peaked around periapsis.
From this expression, we can also obtain the magnitude of the total power radiation by integrating:
\begin{eqnarray}
    \Delta E=\int^{\infty}_{-\infty}P(t) dt=-\int^{\infty}_{-\infty}\dfrac{dE}{dt}\dfrac{dt}{d\xi}d\xi=\dfrac{85  \pi M^{5/2}\mu^2}{12\sqrt{2}r^{7/2}_{\rm min}},\nonumber \\
    \label{total energy quadrupole}
\end{eqnarray}
which exactly matches with the expression given in Ref. \cite{Berry:2010gt}. We have also calculated the total energy in the frequency space, and it matches with the results in Ref. \cite{Berry:2010gt} within the relative error of $0.16\%$. Similar to the energy radiation, we can also obtain the radiation due to angular momentum. By using the parameterization Eq. \eqref{eq:parameterization}, we obtain the instantaneous change of angular momentum as:
\begin{equation}
    \label{eq:angmomeqn}
    \frac{dL_{z}}{dt}=-\frac{2G}{5c^5}\epsilon^{ikl} 
    \ddot{Q}_{ka} \dddot{Q}_{la} = - \frac{48 \sqrt{2} M^{5/2} \mu^2}{5 r^{7/2}_{min} (1 + \xi^2)^5}
\end{equation}
The magnitude of total angular momentum flux radiated is given by:
\begin{eqnarray}
    \Delta L=-\int^{\infty}_{-\infty}\frac{dL_{z}}{dt} dt=-\int^{\infty}_{-\infty}\dfrac{dL_{z}}{dt}\dfrac{dt}{d\xi}d\xi=\frac{6 \pi \mu^2 M^2}{r^2_{min}}.
\end{eqnarray}
%
We will be using the above expression in future sections.  
\subsubsection{Using field theory calculations:}
\begin{figure*}[htp]
  \centering  \includegraphics[width=0.54\textwidth]{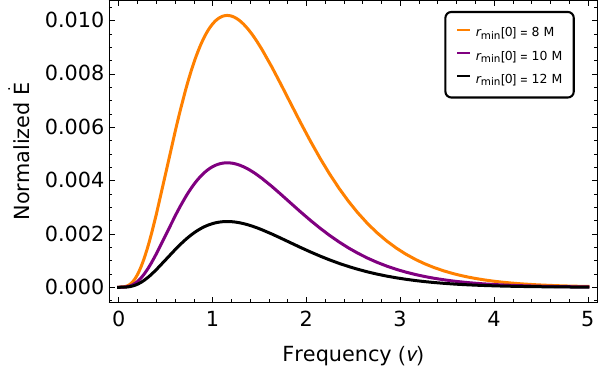} 
  \caption{Power Spectrum for different $r_{\rm min}$ values in frequency space.}
  \label{fig:Power spectrum freq}
\end{figure*}

The other interesting aspect to check is whether the field theory calculation for $\Delta E$ provides a similar result as obtained from the quadrupole formula. By using the expression for $E_{\rm gw}$ given in Eq. \eqref{energy radiated field theory}, we obtain:
\begin{eqnarray}
    E_{gw} = \frac{8}{5} \sum_{\omega'_{n}}\left(T_{ij}(\omega_n') T^*_{ji}(\omega_n') - \frac{1}{3}|T^i_{ i}(\omega_n')|^2 \right) \omega'^2_{n}, 
    \label{EGW in omega}
\end{eqnarray}
where we have used the following relation
\begin{equation}
    \int \omega^2 \delta(\omega'_{n} - \omega)d\omega = \omega'^2_n.
    \label{eq:omegaprime}
\end{equation}
From Eqs. \eqref{eq:Txyfinal}-\eqref{eq:Txxfinal}, we had already obtained the Fourier space $T_{ij}$ in terms of $\nu$. Changing Eq. \eqref{EGW in omega} in terms of $\nu$, and the summation to an integral, we obtain (using $\omega'_{n} = \nu \omega_{0}$),
\begin{equation}
        E_{gw} = \frac{8 \omega^3_0}{5} \int_{\nu}\left(T_{ij}(\nu) T^*_{ji}(\nu) - \frac{1}{3}|T^i_{ i}(\nu)|^2 \right) \nu^2 d\nu
        \label{E integrall},
\end{equation}
and it is shown in Fig. (\ref{fig:Power spectrum freq}). It can be easily noticed that as $\nu \rightarrow 0$, the power vanishes, hinting at the fact that there exists no linear memory. Note that the expression of power spectrum in the frequency domain can be directly implemented to obtain the \textit{stochastic GW background} \cite{Allen:1996} produced by the parabolic encounters. This can be particularly useful in the context of Pulsar Timing Arrays \cite{Burke-Spolaor:2018bvk, Tomson:2025ixn}.   

Upon integrating over all $\nu$'s using integrals of Airy functions and their derivatives (c.f. Appendix \ref{sec:Airy}) and further using Parseval's theorem, we finally obtain the energy radiated from the frequency space approach:
\begin{equation}
    \Delta E = \frac{85 \pi M^{5/2} \mu^2}{12 \sqrt{2}
 r^{7/2}_{min}},
\end{equation}
which matches exactly with Eq. \eqref{total energy quadrupole}.
\subsection{Timescales for parabolic events}
%
In the typical two-timescale approximation, there exists a prolonged timescale where the radiation takes place. On the other hand, there exists the orbital timescale where the radial distance and phase changes rapidly, but keeping energy and momentum as constants throughout. If we take the eccentric orbit as an example, the radiation reaction timescale should follow
\begin{equation}
    t_{\rm RR} \sim \dfrac{E_{\rm orb}}{\langle\dot{E}\rangle},
\end{equation}
where $E_{\rm orbit} \sim M \mu/a$, $a$ being the semi-major axis, and $\langle\dot{E}\rangle$ is the average energy radiation \cite{PetersMathews:1963}. Note that for an  equal/nearly-equal mass ratio binary, we can have
\begin{equation}
    \dfrac{t_{\rm RR}}{t_{\rm orbit}} \sim (a/M)^{5/2},
\end{equation}
where we have used, $t_{\rm orbit} \sim a^{3/2}$. This tells us that the radiation timescale is significantly larger compared to the orbital timescale, given a large semi-major axis. However, unlike the bound eccentric orbits, the parabola introduces a sudden jump, similar to a burst -- an impulsive, strong effect. Therefore, it is not like a long-term secular change, and the time-scale analysis becomes subtle. To have a relevant timescale in the parabolic case, we can obtain the ratio of total radiated energy ($\Delta E$) and the instantaneous power radiation ($P$). With this, we obtain:
\begin{equation}
    \dfrac{\Delta E}{P} \sim (r_{\rm min}/M)^{3/2}.
\end{equation}
As one anticipates, this exactly represents the orbital timescale. This means that for a parabolic encounter, both the radiation and orbital timescales are of the same order.

\subsection{Radiation affecting orbital parameters}
\begin{figure*}[htp]
\centering
    \includegraphics[width=1.02\linewidth]{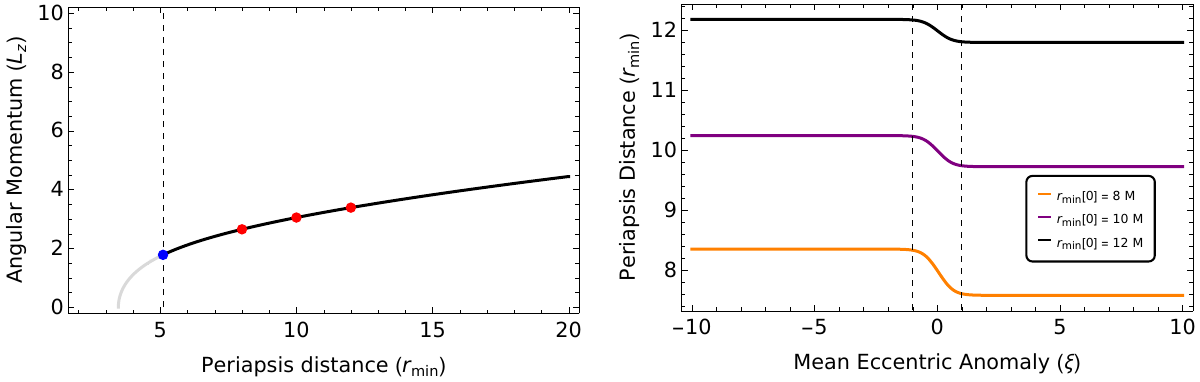}
     \label{rminvslz}
    \caption{In the left panel, the blue point denotes $r_{\rm min} = 5.1 M$, below which, the points are neglected (grey). The black line shows physically relevant cases from which 3 cases ($r_{\rm min} = 8 M, 10 M, 12 M$), as shown with red points, are selected for analysis. The right panel shows the variation of $r_{\rm min}$ with $\xi$ for different instantaneous captures.}
  \label{fig:rmin_subfigures}
\end{figure*}

\begin{figure*}[htp]
\hspace*{-0.1cm}
\includegraphics[scale=.94]{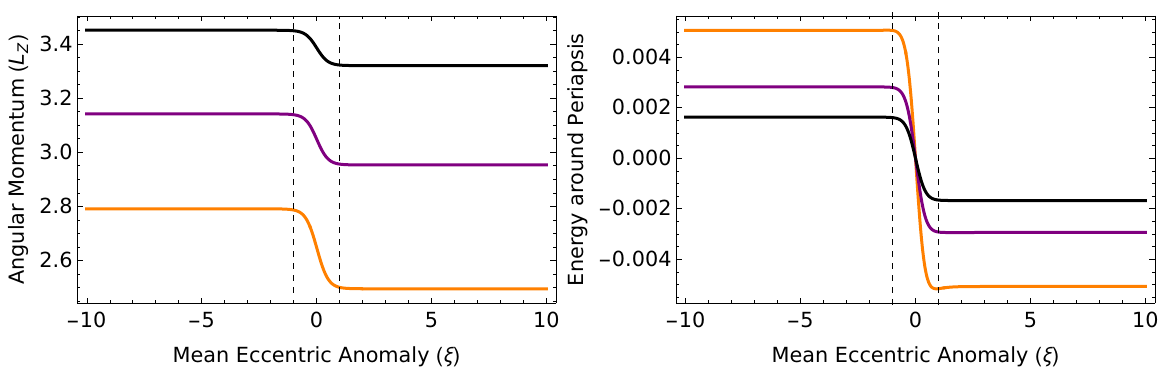}
   \caption{Variation of $L_z$ (left) and $E_{\rm in}$ (right) near the peripasis with respect to $\xi$, respectively, for different instantaneous captures. Different colors in the plot have the same meaning as in Fig. \ref{fig:rmin_subfigures}.}
  \label{fig:Lz_subfigures}
\end{figure*}

Let us now consider the case when the binary starts at $\xi=\xi_0$, radiates energy $\Delta E$ and reaches the periapsis at $\xi=0$ with energy $E_{\rm peri}$. In that case, the following equation is to be satisfied: 
\begin{equation}
    E_{\rm in}=|\Delta E|+E_{\rm peri},
\end{equation}
which indicates that an orbit of initial energy $E_{\rm in}$ radiates energy in the form of GWs, and becomes another orbit with changed energy, i.e., $E_{\rm peri}$. The energy at the periapsis, $E_{\rm peri}$, should follow:
\begin{equation}
\label{eq:total energy}
     E_{\rm peri} = \frac{L_{z}^2}{2 \mu r^2_{min}} - \frac{M \mu}{r_{\rm min}}.
\end{equation}
Now, as the binary initially moves in a parabolic orbit, we should have $E_{\rm in}=0$. Therefore, we always have $E_{\rm peri}<0$, and the orbit becomes bounded and eccentric. It renders the fact that any parabolic orbit undergoing a GW radiation would be captured, and form an elliptical binary. Finally, we can therefore write 
\begin{equation}
    E_{\rm peri}+|\Delta E|=0,
    \label{eq:rmin&Lz_xi0}
\end{equation}
and the above can be readily solved for $r_{\rm min}(\xi=0)$ for a given value of $L_{z}(\xi=0)$. In Fig. (\ref{fig:rmin_subfigures}), the first panel gives a contour plot between $r_{\rm min}$ and $L_{z}$, which represents the points where the above expression is satisfied. Given the radiation is strong near periapsis, we now consider the fact that $r_{\rm min}$ itself starts to change. 
\begin{equation}
    \label{eq:energy - variable Lz}
    \frac{dr_{\rm min}}{d\xi} = \left(\frac{dE}{d\xi} - \frac{L_{z}}{\mu r_{\rm min}^2}\frac{dL_{z}}{d\xi}\right) \left(\frac{M \mu}{2 r^2_{min}} - \frac{L_{z}^2}{\mu r^3_{min}}\right)^{-1}. 
\end{equation}
With the proper parameterization to $\xi$, the change in energy and momentum can be calculated from Eqs. \eqref{eq:power equation} and \eqref{eq:angmomeqn}. Once substituted, we can solve the above expression along with 
\begin{eqnarray}
    \dfrac{dL_z}{d\xi}=\dfrac{dL_z}{dt}\dfrac{dt}{d\xi},
\end{eqnarray}
and solve for $r_{\rm min}(\xi)$ and $L_{z}(\xi)$. Note that the initial conditions are to be calculated from Eq. \eqref{eq:rmin&Lz_xi0}. In the second panel of Fig. \eqref{fig:rmin_subfigures}, the change of $r_{\rm min}$ is shown with changing $\xi$, as the binary passes through the periapsis. As expected, the instantaneous change in $E$ and $L_z$ results in a finite jump in $r_{\rm min}$ near perapsis, signifying the capture. For completeness, Fig. \eqref{fig:Lz_subfigures} shows the $\xi$ variation of the angular momentum and energy as well, both showing similar finite jumps near periapsis. 
%
\par
Finally, we end on a subtle note here about the parameter space, the points left of the blue one in the first panel of Fig. \eqref{fig:rmin_subfigures} are deemed non-physical. This is because with those initial values of $r_{\rm min}$ and $L_{z}$, after the jump, the resulting $r_{\rm min}$ falls below the Schwarzschild radius, which implies head-on collisions between binary constituents. We do not consider these situations in our analysis. The vertical dashed lines depict the nearby region around the periapsis $(\xi = 0)$ where the burst takes place. 

\section{Non-Linear Memory and jumps \label{sec:nonlinmem}}
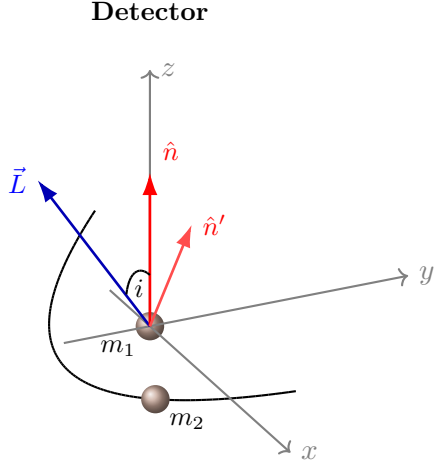
\begin{figure}
\centering
\tdplotsetmaincoords{65}{65}
\def\i{-20} 
%
\definecolor{warm1}{RGB}{206,137,116}
\definecolor{warm2}{RGB}{197,164,149}

\begin{tikzpicture}[tdplot_main_coords, scale=1.25,
  arrow/.style={-{Latex[length=3mm,width=2mm]}, line width=1pt},
  thinline/.style={line width=0.6pt, gray!70}
]

\fill[white!10] (2,-4) rectangle (-6,6.2);
\def\b{0.25}
\def\a{1.0}
\def\xmin{-2.5}
\def\xmax{2.5}
\def\dx{0.05}

\foreach \iindex in {0,...,99}{
  \pgfmathsetmacro{\x}{\xmin + \iindex*\dx}
  \pgfmathsetmacro{\nx}{\x + \dx}
  \pgfmathsetmacro{\yflat}{\b*(\x*\x) - \a}
  \pgfmathsetmacro{\nyflat}{\b*(\nx*\nx) - \a}
  \pgfmathsetmacro{\yrot}{\yflat*cos(\i)}
  \pgfmathsetmacro{\zrot}{-\yflat*sin(\i)}
  \pgfmathsetmacro{\nyrot}{\nyflat*cos(\i)}
  \pgfmathsetmacro{\nzrot}{-\nyflat*sin(\i)}
  \draw[thick, black] (\x,\yrot,\zrot) -- (\nx,\nyrot,\nzrot);
}

\coordinate (m1) at (0,0,0); 
\pgfmathsetmacro{\ytmp}{(\b*(1.3*1.3) - \a)*cos(\i)}
\pgfmathsetmacro{\ztmp}{-(\b*(1.3*1.3) - \a)*sin(\i)}
\coordinate (m2) at (1.3,\ytmp,\ztmp);
\coordinate (D)  at (0,0,3);

\shade[ball color=warm2] (m1) circle (0.15cm);
\shade[ball color=warm2] (m2) circle (0.15cm);
\node[below left=2pt of m1, font=\small] {$m_1$};
\node[below right=2pt of m2, font=\small] {$m_2$};

\node[above=15 pt of D, font=\bfseries\small] {Detector};

\draw[->, thick, color=gray] (-1,0,0) -- (3.5,0,0) node[right]{$x$};
\draw[->, thick, color=gray] (0,-1,0) -- (0,3,0) node[right]{$y$};
\draw[->, thick, color=gray] (0,0,0) -- (0,0,3) node[right]{$z$};

\draw[arrow, blue!70!black] (m1) -- ++(-1.3,-0.7,1.3)
  node[left, font=\small, color=blue] {$\vec{L}$};

\draw[arrow, red] (m1) -- ++(0,0,1.8)
  node[above right=1pt, font=\small, color=red] {$\hat{n}$};

\draw[arrow, red!70] (m1) -- ++(0.4,0.3,1.3)
  node[right, font=\small, color=red] {$\hat{n}'$};

\begin{scope}[shift={(m1)}]
  \def\r{0.6}
  \pgfmathsetmacro{\dang}{6.5}
  \pgfmathsetmacro{\angmin}{0}
  \pgfmathsetmacro{\angmax}{60}
  \pgfmathsetmacro{\prevx}{-\r*sin(\angmin)}
  \pgfmathsetmacro{\prevz}{\r*cos(\angmin)}
  \foreach \k in {1,...,12}{
    \pgfmathsetmacro{\ang}{\angmin + \k*\dang}
    \pgfmathsetmacro{\xx}{-\r*sin(\ang)}
    \pgfmathsetmacro{\zz}{\r*cos(\ang)}
    \draw[thick] ({\prevx},0,{\prevz}) -- ({\xx},0,{\zz});
    \xdef\prevx{\xx}
    \xdef\prevz{\zz}
  }
  \node at ({-\r*0.45},0,{ \r*0.55}) [font=\small] {$i$};
\end{scope}

\end{tikzpicture}
\caption{A cartoon of the binary lying on the $x-y$ plane, oriented at an angle $i$ with respect to the observer/detector.}
\label{fig:orbit_diagram}
\end{figure}

We then proceed to the other key objective of this paper -- focusing on non-linear memory! As of now, our picture is clear, the source emits radiation violently near the periapsis, with the radiation dying with a tail at infinity, while the detector slowly comes back to its original position. Following the discussion in section-\eqref{sec:effectiveft}, we now seek to obtain the transverse-traceless wave function responsible for the non-linear memory signal for our binary undergoing a parabolic encounter using Eq. \eqref{eq:nonlinear wavefunction}. To proceed further, recall Eq. \eqref{energy radiated field theory} and \eqref{eq:omegaprime}, we obtain the energy radiated in the direction $d\Omega'$ expressed by the formula:
\begin{equation}
\label{eq:radiated_energy}
    \frac{dE_{gw}}{dt' d\Omega'} = \frac{\kappa^2}{8}\sum_{\nu = 0}^{\infty} T_{ij} T_{kl}^* \Lambda_{ij, kl}(\hat{n}') \frac{\omega'^{2}}{(2\pi)^2},
\end{equation}
where, $\omega' = \nu \omega_0$ are the harmonics as before. The stress energy tensor $T_{ij}$ can be written in the matrix form as:
\begin{equation}
    \mathrm{T} = \frac{\pi r_{\rm min}^2}{\tau} \begin{pmatrix}
        q_{xx} & iq_{xy} & 0\\
        i q_{xy} &  q_{yy} & 0 \\
        0 & 0 & 0 \\
    \end{pmatrix},
\end{equation}
with, $i~\text{and}~j$ represents $(x,y,z)$ values, and
\begin{align}
q_{xx} &= 2^{7/3} \nu^{1/3} Ai'((2 \nu)^{2/3}) + 2^{2/3} \nu^{-1/3} Ai((2 \nu)^{2/3}),\\ \nonumber
q_{xy} &= -2^{8/3} \nu^{2/3} Ai((2 \nu)^{2/3}),~
q_{yy} = -2^{7/3} \nu^{1/3} Ai'((2 \nu)^{2/3}).
\end{align}
Note that here $\tau$ is the characteristic timescale of the encounter. The structure of $\mathrm T$ signifies that we are dealing with a planar source as in the preceding cases, and $z$ components do not contribute. But, in general, a detector can be placed at any inclination with this plane of the parabola (see Fig.\eqref{fig:orbit_diagram}). 
\par
Let us consider such a case now, where the axis of rotation of the binary $\Vec{L}$ makes an angle $i$ with the $z-$axis where the direction vector is given by $\hat{n}' = (\sin \theta' \cos \phi', \sin \theta' \sin \phi', \cos \theta')$. For the rotated system, the stress energy tensor is given by a similarity transformation: $\mathrm{T'} = \mathrm{R}\mathrm{T} \mathrm{R^{\mathrm{T}}}$, where $\mathrm{R}$ is a rotation matrix around the $x$ direction.
Under this transformation $\mathrm{T'}$ can be rewritten as: 
\begin{equation}
    \mathrm{T}' = \frac{\pi r_{\rm min}^2}{\tau}  
    \begin{pmatrix}
        q_{xx} & iq_{xy} \cos i & i q_{xy} \sin i \\
        i q_{xy} \cos i &  q_{yy} \cos^2 i & q_{yy} \cos i \sin i \\
        i q_{xy} \sin i & q_{yy} \cos i \sin i & q_{yy} \sin^2 i \\
    \end{pmatrix}
\end{equation}
Since both $\mathrm{T}$ and $\mathrm{T'}$ aren't traceless, the matrix product $T_{ij}T_{kl}^* \Lambda_{ij, kl}(\hat{n}')$ is written as a sum over matrix products as shown in Ref. \cite{Hait:2022ukn}. The matrix product $T_{ij}T_{kl}^* \Lambda_{ij, kl}(\hat{n}')$ in the parabolic case therefore, can be simply expressed as a function of $\nu$, $\theta'$ and $\phi'$, 
\begin{equation}
\label{eq:interaction_term}
        T_{ij}T_{kl}^* \Lambda_{ij, kl}(\hat{n}') = \frac{r_{\rm min}^4 \pi^2 }{\tau^2}~ \mathrm{I}(\nu, \theta', \phi').
\end{equation}
The exact expression of $\mathrm{I}(\nu, \theta', \phi')$ is not physically illuminating for our particular case of parabolic encounter, and contains polynomials of $q_{\alpha\beta}$ and trigonometric functions of $i$. We will refrain from mentioning it explicitly here and relegate it to Appendix \ref{sec:IFunc}. We then substitute Eq. \eqref{eq:interaction_term} into Eq. \eqref{eq:radiated_energy} which gives us the rate of energy radiated as
\begin{equation}
\label{eq:dedt'domega'}
    \frac{dE_{gw}}{dt'd\Omega'} = \frac{\pi \nu^2}{r^2_{\rm min}}\mathrm{I}(\nu, \theta', \phi').
\end{equation}
Furthermore, in order to obtain the transverse-traceless wave function, we must evaluate the angular integral in Eq. \eqref{eq:nonlinear wavefunction} over $\theta'$ and $\phi'$ and then sum over $\nu$ orders of the Airy functions in Eq. \eqref{eq:radiated_energy}. The angular integrals yield
\begin{equation}
\begin{split}
\label{eq: Axx}
    [A_{xx}]^{TT} = \sum_{\nu = 0}^{\infty} \nu^2 \int^{\pi}_0 \sin \theta' d\theta' \int^{2\pi}_0 d\phi' ~\mathrm{I}(\nu, \theta', \phi') \times \nonumber\\
     \hspace{4cm}\frac{1}{2}(1 + \cos \theta') \cos 2\phi',\\ 
    = \frac{\pi}{16384}\left(-4615 + 5292 \cos 2i + 91 \cos 4i \right),
\end{split}
\end{equation}
and similarly,
\begin{align}
\label{eq:Ayy}
    [A_{xy}]^{TT} = \sum_{\nu = 0}^{\infty} \nu^2 \int^{\pi}_0 \sin \theta' d\theta' \int^{2\pi}_0 d\phi' ~ \mathrm{I}(\nu, \theta', \phi') \times \nonumber \\
    \frac{1}{2}(1 + \cos \theta') \sin 2\phi' = 0.
\end{align}
The other components can be obtained using the relations $[A_{yy}]^{TT} = -[A_{xx}]^{TT}$ and $[A_{xy}]^{TT} = [A_{yx}]^{TT}$. Note that all of these $\nu$ sums are smooth ones. Substituting Eqs. \eqref{eq:dedt'domega'}-\eqref{eq:Ayy} into the expression for the non-linear memory waveform in Eq. \eqref{eq:nonlinear wavefunction} gives,
\begin{equation} 
\label{eq:hcomponents}
\begin{aligned}
    \left[h_{xx}^{mem} \right]^{TT} &= \frac{4 \pi}{r} [A_{xx}]^{TT} \int^{T_r}_{-\infty} dt' \frac{1}{r_{\rm min}^2}, \\
    \left[h_{yy}^{mem} \right]^{TT} &= \frac{4 \pi}{r} [A_{yy}]^{TT} \int^{T_r}_{-\infty} dt' \frac{1}{r_{\rm min}^2}, \\ 
     \left[h_{xy}^{mem} \right]^{TT} &= \left[h_{yx}^{mem} \right]^{TT} = 0,
\end{aligned}   
\end{equation}
where $T_r$ is the retarded time. The temporal integrals above can be suggestively rewritten in terms of $\xi$ as follows:
\begin{equation}
    \int^{\epsilon_0}_{-\epsilon_0} \dfrac{dt'}{d\xi} \frac{1}{r_{\rm min}^2}d\xi=\int f(\xi)d\xi.
\end{equation}
where $f(\xi)$ is the integrand. Notice that, as we are modeling a burst-like encounter, we adjusted the limits of the integration accordingly. We assume, motivated by physics considerations, that the encounter takes place within an infinitesimal interval $(-\epsilon_0,\epsilon_0)$ around the periapsis approach itself. This is also consistent with Eq. \eqref{eq:power}, which suggests that for large $\xi$, the power falls as $\xi^{-10}$, and therefore, only the short passage in $\xi$ will contribute.
\par
Let us now try to understand the structure of these waveforms heuristically, since it depends on the exact details of $r_{\rm min}$. In the above expression, we notice that $r_{\rm min}$ can be fitted with a hyperbolic function as the radial separation near periapsis transitions sharply from one asymptotic regime to another, hence we can write:
\begin{equation}
    \dfrac{1}{r_{\rm min}^2} \sim a+2b \tanh(\xi),
    \label{eq:rmin}
\end{equation}
where $a$ and $b$ are two fitting parameters. Note that the motivation for using the hyperbolic tangent as the fitting function is that it represents the simplest smooth function that connects two asymptotic plateaus through a monotonic transition. A more rigorous, first-principle derivation of the precise functional form would require a detailed treatment of the full non-linear dynamics. Nevertheless, we believe that this simple analytic ansatz captures the essential qualitative features of the signal. We can now substitute the above expression into Eq. \eqref{eq:hcomponents}, and obtain
\begin{equation}
    h \sim \dfrac{4 \pi}{r}[A]^{TT} \int^{\epsilon_0}_{-\epsilon_0} f(\xi)d\xi.
\end{equation}
Note again that we are only considering the short passage of time within which the jump takes place, say $\xi \in (-\epsilon_0,\epsilon_0)$.
Given that $\epsilon_0$ should be small for detectable events \cite{Mukherjee:2020hnm}, we may assume that $f(\xi)$ behaves as a straight line around $\xi=0$. Therefore, we can approximate $f(\xi)$ as
\begin{equation}
    f(\xi)\approx 2\tau(1+\xi^2)(a+2b~\xi).
\end{equation}
Therefore, the expression for the perturbation around $\xi=0$ would produce
\begin{equation}
    h \sim 4 \pi[A]^{TT}~ \dfrac{2\tau}{r}( a\xi + b\xi^2 +  a\xi^3/3+ b\xi^4/2 + \mathcal{O}(\xi^5))\Big|^{\epsilon_0}_{-\epsilon_0}.
\end{equation}
Hence, the jump in  perturbation is given as
\begin{equation}
    \Delta h \sim \dfrac{8\pi \tau}{r}[A]^{TT} a\epsilon_0 + \mathcal{O}(\epsilon^3_0),
    \label{eqn: delta h final}
\end{equation}
which contains an explicit dependence on the window parameter $\epsilon_0$. Since this parameter is introduced solely to localize the burst around the periapsis, the corresponding amplitude should not be interpreted as a unique physical prediction for the nonlinear memory. The fit only provides a qualitative estimate of the characteristic scale and step-like structure of the memory-like transition. Coming back to Eq. \ref{eqn: delta h final}, the value of $a$ can be fixed from Eq. \eqref{eq:rmin}
\begin{equation}
    a=\lim_{\xi \rightarrow 0} \dfrac{1}{r_{\rm min}^2}.
\end{equation}
Note that $\tau$ varies with $r_{\rm min}$ as in Eq. \eqref{tauscale}, so the total scaling of $\Delta h$ comes out to be $r_{\rm min}^{-1/2}$.
\begin{figure*}[htp]
  \centering
  \includegraphics[width=1.0\linewidth]{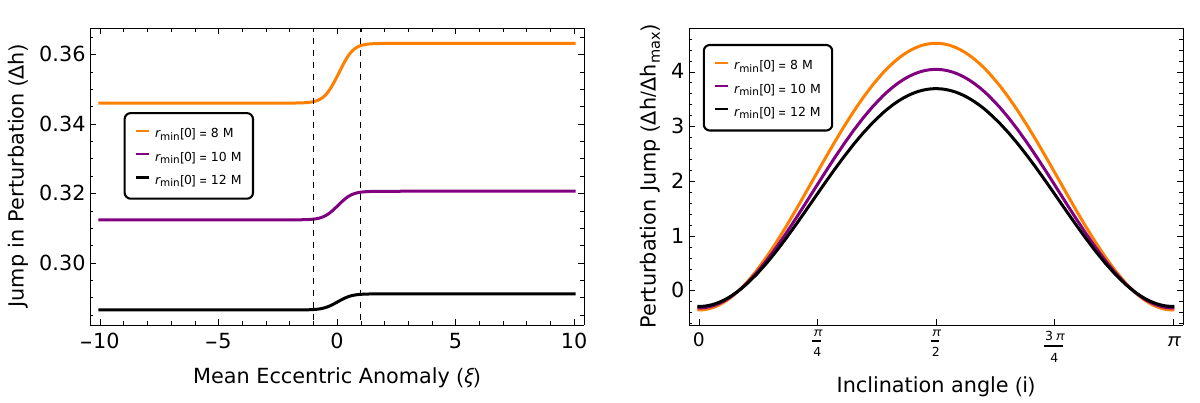}
  \caption{Left panel: Corresponding memory ``jumps" for ($r_{\rm min} , L_{z}$) values for instantaneous captures. Right panel: Strain amplitude dependence on the inclination angle of the binary (i).}
  \label{fig:deltah}
\end{figure*}
\par
For this structure, variation of memory waveforms Eq. \eqref{eqn: delta h final} both with changing $r_{\rm min}$ at periapsis values and the inclination angle $i$ has been presented in Fig. \eqref{fig:deltah}. For all values of $r_{\rm min}$ above the Schwarzschild threshold, we get a finite value of the jump around periapsis. In the right panel of Fig. \eqref{fig:deltah}, one can see that the strain is maximized for face-on orientations (and minimized for edge-on orientations), which is consistent with the projection effects of the quadrupole radiation. 
\section{Discussions and Conclusion} \label{section6}
In the present work, we discussed an interesting limit of a binary system -- the parabolic encounter. The key objective of the paper is to work out the GW memory calculations from a field-theoretic frequency space approach. We started out by motivating the problem from the various subtleties associated with these encounters, both from the limiting point of view and from the universal soft theorem understanding. Working with a particular intrinsic parameterization, we worked out the graviton emission amplitudes using the classical stress tensor associated with the radiation. While linear memory associated with the binary can simply be obtained to be zero from the (time domain) soft factor associated with particles taking part in the process, the non-linear or graviton-graviton interaction part is much more intricate. We used the field theoretical formula for secondary radiation that directly gives rise to the memory waveform in the frequency space for the latter.
\par
In many ways, the parabolic encounter is indeed special. Physically speaking, the symmetry in the velocity asymptotes directly tells us that there should be no linear memory attached to the system. However, we have noticed that for events following a parabolic trajectory, there is a sharp change of stress energy tensor in the $\nu \rightarrow 0$ limit, as evident by the pole of $\nu^{-1/3}$ we encountered there. Mathematically, this appears as a consequence of how Airy functions behave close to $\nu \rightarrow 0$, which distinguishes the parabolic encounter from eccentric or hyperbolic!  It is more like a transient effect which dominates at the low frequency range; however, it does not produce any permanent change or memory. Non-linear memory, on the other hand, depends on graviton back-reaction, which would be present even for a short and powerful burst of radiation that occurs in this case near periapsis. The power spectrum of the energy emitted during such parabolic encounters is relevant to probe the contribution of such encounters to the stochastic gravitational wave power spectrum using pulsar timing arrays. The inevitable change in energy and angular momentum due to these close encounters will mean that the closest point of approach in a binary will see a sudden jump scale as $r_{\rm min}^{-1/2}$, and will be captured as a bound orbit as the total energy dips below zero. We followed through with this logic and worked out how this burst of radiation changes all the binary parameters. It is physically instructive to think of the sudden change in the closest approach around the periapsis as a sharp, localized step, which leads to the associated jump in the memory waveform. 
\par
Let us now briefly comment on the observational implications of our results. As far as the linear memory from binaries is concerned, both hyperbolic and eccentric limits are studied extensively \cite{Hait:2022ukn, de2016gravitational}. For hyperbolic encounters with large eccentricity ($ e \gg 1$), the asymptotic velocities of the binary components are significant, leading to a considerable linear memory contribution. However, such encounters involve weak interactions and therefore, a relatively weaker GW signal. If we now approach the parabolic limit ($e \rightarrow 1^{+}$), the nature of the encounter changes. The interaction near periapsis becomes stronger, which leads to a strong GW signal, but at the cost of decreased asymptotic velocities. This causes the linear memory to suppress and eventually disappear for the exact parabolic case. Therefore, there exists an interesting trade-off -- a weaker GW signal with larger linear memory vs. a stronger GW signal with weaker linear memory. With these considerations, we believe that near-parabolic encounters are better identified with their burst-like signatures, and not with linear memory alone. From its frequency-domain counterpart, parabolic events contain a unique pole structure, and this sets them apart from even near parabolic orbits. As we have already discussed, this particular effect may be visible in the early-time portion of the tail, right after the burst at periapsis.  
\par
To summarize, the parabolic encounter emerges as a physically intriguing limiting case of binary events. On one hand, it introduces nontrivial features at the low frequency range even without contributing to any linear memory. On the other hand, the burst-like feature close to the periapsis modifies the trajectory for a short span of time and contributes to non-linear memory. Even if the present GW detectors may not be ideal for probing GW signals with memory (see Ref. \cite{PhysRevLett.117.061102} to detect GW memory with present detectors), future GW detectors with enhanced sensitivity may have a better scope to study these events \cite{Ghosh_2023, PhysRevD.107.044023}. Our work clarifies the theoretical foundations for studying these encounters, and perhaps can motivate more investigations along these avenues in future. 

\acknowledgments
The authors would like to thank Shaunak Patharkar for discussions and comments on the manuscript. S.D. acknowledges the financial support from BITS Pilani through the seed grant NFSG/PIL/2023/P3794 for the duration of this work. He (S.D.) is currently supported by a doctoral fellowship (88887.191515/2025-00) from Coordenação de Aperfeiçoamento de Pessoal de Nível Superior - CAPES, Brazil. A.C. acknowledges BITS Pilani, India, for research facilities and financial support for carrying out this work. ABan is supported in part by an OPERA grant and a seed grant NFSG/PIL/2023/P3816 from BITS-Pilani, and further an early career research grant ANRF/ECRG/2024/002604/PMS from ANRF India. He also acknowledges financial support from the Asia Pacific Center for Theoretical Physics (APCTP) via an Associate Fellowship. S. Mukherjee is thankful to the Inspire Faculty Grant (DST/INSPIRE/04/2020/001332) from DST, Govt. of India, Prime Minister Early Career Research Grant (ANRF/ECRG/2024/004108/PMS) by ANRF, Govt. of India, and the New Faculty Seed Grant (NFSG/PIL/2023/P3794) provided by BITS Pilani (Pilani), India, for financial support. He (S. Mukherjee) is also grateful to the Visiting Associateship Program at IUCAA, Pune, for academic visits where a part of this work was carried out.

\appendix
\section{Airy Integrals} \label{sec:Airy}
For real values of $z$, the Airy function of the first kind $Ai(z)$ and its derivative (denoted by primes and with respect to the argument $z$) are defined through simple integral representations as follows \cite{Watson:1944}:
\begin{equation}
\label{eq:Airy_Integral_Def}
\begin{split}
        Ai(z) = \displaystyle \frac{1}{\pi}\int^{\infty}_{0} \cos \left(\frac{\xi^3}{3} + z \xi \right) d\xi,\nonumber \\
        Ai'(z) = \displaystyle -\frac{1}{\pi}\int^{\infty}_{0} \xi \sin \left(\frac{\xi^3}{3} + z \xi \right) d\xi.
\end{split}
\end{equation}
The derivatives of the Airy function $Ai(z)$ and its derivative $Ai'(z)$ can also be simply expressed in terms of Airy functions as follows:
\begin{equation}
\label{eq:Airy_Derivative_Def}
    \begin{split}
        \displaystyle \frac{\partial Ai(z)}{\partial z} = Ai'(z), ~~~~
        \displaystyle \frac{\partial Ai'(z)}{\partial z} = z Ai(z).
    \end{split}
\end{equation}
Therefore, by extension: 
\begin{equation}
\label{eq:Airy_Higher_Derivative}
    \frac{\partial^2 Ai'(z)}{\partial z^2} = Ai(z) + z Ai'(z).
\end{equation}
Using these properties of Airy functions as shown in Eqs. \eqref{eq:Airy_Integral_Def} to \eqref{eq:Airy_Higher_Derivative}, we solve the integrals in Eq. \eqref{eq:Txx-unsolved} and Eq. \eqref{eq:Tyy-unsolved} to obtain:
\begin{align}
    \int^{\infty}_{-\infty} \xi e^{2i \nu \left(\frac{\xi^3}{3} + \xi \right)} d\xi &= -2 i \pi (2 \nu)^{-2/3} Ai'((2 \nu)^{2/3}),\\
    \int^{\infty}_{-\infty} \xi^2 e^{2i \nu \left(\frac{\xi^3}{3} + \xi \right)} d\xi &= -2\pi (2 \nu)^{-1/3} Ai((2 \nu)^{2/3}),\\
    \int^{\infty}_{-\infty} \xi (1 - \xi^2) e^{2i \nu \left(\frac{\xi^3}{3} + \xi \right)} d\xi &= -2 \pi i \Big[ 2 (2 \nu)^{-2/3} Ai'((2 \nu)^{2/3}) \nonumber \\
    & +(2\nu)^{-4/3} Ai((2 \nu)^{2/3}) \Big].
\end{align}
The above equations are used to obtain the stress-energy tensor components in terms of the Airy function and its derivative Eqs. \eqref{eq:Txyfinal} to \eqref{eq:Txxfinal}. At the Zero Frequency Limit (ZFL), when $\nu \rightarrow 0$, the expansions of the Airy function and its derivative go as:
\begin{align}
\label{eq:AiryLim}
    Ai((2\nu)^{2/3}) &\simeq \frac{1}{3^{2/3} \Gamma(2/3)} - \frac{2^{2/3} \nu^{2/3}}{3^{1/3} \Gamma(1/3)} + \frac{2 \nu^2}{3^{11/3} \Gamma(2/3)}  \nonumber \\
    & \hspace{5cm} +\mathcal{O}(\nu^{7/3}),
\end{align}
\begin{align}
\label{eq:AiryDLim}
    Ai'((2\nu)^{2/3}) &\simeq -\frac{1}{3^{1/3} \Gamma(1/3)} + \frac{2^{1/3} \nu^{4/3}}{3^{2/3} \Gamma(2/3)} - \frac{4 \nu^2}{3^{4/3} \Gamma(1/3)}  \nonumber \\
   & \hspace{5cm} +\mathcal{O}(\nu^{7/3}), \nonumber \\
\end{align}
where, $\Gamma(z)$ is the gamma function commonly defined as
\begin{equation}
    \Gamma(z) = \int^{\infty}_0 t^{z - 1} e^{-t} dt.
\end{equation}
The above expansion of the Airy function and its derivatives as $\nu \rightarrow 0$ indicates the absence of $\frac{1}{\nu}$ poles in Eqs. \eqref{eq:Txyfinal} to \eqref{eq:Txxfinal}. Moreover, the calculation of the total power radiated using Eq. \eqref{energy radiated field theory} requires one to use integrals of bilinears of Airy function and derivatives thereof \cite{DeVittori:2012da}. The following integrals have been used to evaluate Eq. \eqref{E integrall}:
\begin{align}
    \int^{\infty}_{0} \nu^{4/3}[Ai((2\nu)^{2/3})]^2 d\nu = \frac{5}{512}\frac{1}{2^{1/3}},
    \\
    \int^{\infty}_{0} \nu^{10/3} [Ai((2\nu)^{2/3})]^2 d\nu = \frac{1155}{131072}\frac{1}{2^{1/3}},
    \\
    \int^{\infty}_{0} \nu^2 Ai((2\nu)^{2/3}) Ai'((2\nu)^{2/3}) d\nu = -\frac{35}{4096},
    \\
    \int^{\infty}_{0} \nu^{8/3} [Ai'((2\nu)^{2/3})]^2 d\nu = \frac{1365}{65536}\frac{1}{2^{2/3}} 
\end{align}
All of which can be verified using standard integrals. 
\section{A limit from hyperbolic to parabolic}\label{sec:appB}
Let us try to understand the fractional pole in Eq. \eqref{fracpole} better by physically motivating it from a consistently taken limit of the gravitational waveform of the hyperbolic case. The parameterization in this case is
\begin{eqnarray}
   & x(\xi)=a(e-\cosh\xi),~~ y(\xi)=b \sinh\xi,\nonumber \\
   & z(\xi)=0,~~\omega^{\prime}t/\nu=\omega_0 t=(e \sinh\xi-\xi).
\end{eqnarray}
The limit we want to take on the gravitational radiation here is subtle, as it concerns taking both a ZFL and a $e\to 1$ limit \cite{de2016gravitational}. The double scaling associated with these two may need to be handled particularly. Just to be as clear as possible, let us write the Fourier transform of the gravitational wave strain in a suggestive manner:
\begin{equation}
\tilde{h}_{ij}(\omega) = \frac{4G}{r} \int_{-\infty}^{\infty} dt\, e^{i\omega t} \ddot{Q}_{ij}(t).
\label{eq:fourier_strain}
\end{equation}
After integration by parts and a change of variables to $\xi$, this becomes:
\begin{equation}
\tilde{h}_{ij}(\omega) = \frac{4G}{r} \int_{-\infty}^{\infty} d\xi\, F_{ij}(\xi)\, e^{i\omega t(\xi)},
\label{eq:strain_xi}
\end{equation}
where $F_{ij}(\xi) = \frac{d}{d\xi}\left(\frac{dQ_{ij}}{d\xi} \frac{d\xi}{dt}\right)$ is a smooth function, and dynamical details are contained in the phase which contributes dominantly when the radiation burst happens near $\xi = 0$, i.e. the periapsis.
Let $e = 1 + \varepsilon$ with $0 < \varepsilon \ll 1$. For fixed periapsis distance $r_{\rm min}$, we have:
\begin{equation}
r_{\rm min} = a(e - 1) = a\varepsilon \quad \Rightarrow \quad a = \frac{r_{\rm min}}{\varepsilon}
\label{eq:periapsis_relation}
\end{equation}
The orbital frequency thus scales as:
\begin{equation}
\omega_0 = \sqrt{\frac{GM}{a^3}} = \sqrt{\frac{GM \varepsilon^3}{r_{\rm min}^3}} \propto \varepsilon^{3/2}.
\end{equation}
For fixed reference frequency $\omega'$, the ZFL parameter diverges:
\begin{equation}
\nu = \frac{\omega'}{\omega_0} \propto \omega' \varepsilon^{-3/2} \to \infty \quad \text{as} \quad \varepsilon \to 0.
\label{eq:nu_divergence}
\end{equation}
This divergence indicates the incompatibility of the naive parabolic limit with the standard ZFL procedure, as we discussed before. Now, to be more careful and precise, we expand the time function around $\xi = 0$:
\begin{align}
e \sinh\xi - \xi &= (1 + \varepsilon)\left(\xi + \frac{\xi^3}{6} + \frac{\xi^5}{120} + \cdots\right) - \xi \nonumber \\
&= \varepsilon\xi + \frac{\xi^3}{6} + \varepsilon\frac{\xi^3}{6} + \frac{\xi^5}{120} + \cdots
\label{eq:time_expansion_full}
\end{align}
To leading order in $\varepsilon$ and $\xi$ the phase becomes:
\begin{equation}
\omega t(\xi) \approx \nu\left(\varepsilon\xi + \frac{\xi^3}{6}\right).
\label{eq:phase_approx}
\end{equation}
The character of the integral in Eq. \eqref{eq:strain_xi} depends on which term in the phase given above dominates. Let $\xi_*$ be the typical scale of the stationary-phase region. The cubic term is of the order $\mathcal{O}(\nu \xi_*^3)$. For this region to contribute to the integral, we require $\nu \xi_*^3 \sim 1 \quad \Rightarrow \quad \xi_* \sim \nu^{-1/3}$. Now examine the linear term in this region:
\begin{equation}
\nu \varepsilon \xi_* \sim \nu \varepsilon \nu^{-1/3} = \varepsilon \nu^{2/3}.
\label{eq:linear_term_scaling}
\end{equation}
For the cubic term to dominate the physics (giving the parabolic behavior rather than hyperbolic), we require an extra condition on the double limit, which lets us access the Airy regime:
\begin{equation}
\varepsilon \nu^{2/3} \ll 1 \implies \varepsilon \ll \nu^{-2/3}.
\label{eq:dominance_condition}
\end{equation}
So precisely, one has to take the $\nu\to 0$ while scaling the $\varepsilon$ to zero faster than how $\nu^{2/3}$ does. Note that this is a very precise window. 
One should further note here that for $\varepsilon\nu^{2/3} \gg 1$, the linear term will dominate and the linear phase will Fourier transform into a delta function and cutting off the integral near periapsis, we will recover the standard hyperbolic scaling for the signal. At the transitional regime $ \varepsilon\nu^{2/3}\sim 1$, things are a bit more interesting, and we can get both kinds of scalings depending on the arguments.
\par
Under the double limit condition \eqref{eq:dominance_condition} that tips the scales towards parabolic orbits, the linear term in the phase is negligible, and we have:
\begin{equation}
\omega t(\xi) \approx \frac{\nu}{6} \xi^3.
\end{equation}
The strain integral \eqref{eq:strain_xi} becomes:
\begin{equation}
\tilde{h}_{ij}(\omega) \approx \frac{4G}{r} F_{ij}(0) \int_{-\infty}^{\infty} d\xi\, \exp\left(i \frac{\nu}{6} \xi^3\right),
\label{eq:integral_cubic}
\end{equation}
where we've approximated $F_{ij}(\xi) \approx F_{ij}(0)$ since the main contribution comes from small $\xi$.
Now make the change of variables:
\begin{equation}
y = \left(\frac{\nu}{6}\right)^{1/3} \xi, \quad d\xi = \left(\frac{6}{\nu}\right)^{1/3} dy.
\label{eq:rescaling}
\end{equation}
The integral transforms to:
\begin{equation}
\tilde{h}_{ij}(\omega) \approx \frac{4G}{r} F_{ij}(0) \left(\frac{6}{\nu}\right)^{1/3} \int_{-\infty}^{\infty} dy\, e^{i y^3}.
\label{eq:rescaled_integral}
\end{equation}
Now, we know that the standard Fresnel integral \cite{Gradshteyn:1943cpj}
\begin{equation}
I_0 = \int_{-\infty}^{\infty} dy\, e^{i y^3} = \frac{2\pi}{3^{2/3} \Gamma(2/3)}. 
\label{eq:cubic_integral}
\end{equation}
This is a finite constant. Hence, the $\tilde{h}_{ij}(\omega) $ given this window in the parameter space clearly scales with $\nu^{-1/3}$ in the parabolic regime. This is the exact pole we have seen in our calculations.
\par
To round up our discussion, let us also think a little about what happens away from the stationary wave approximation in this small parameter window. The function $F_{ij}(\xi)$ in this case can be expanded in a Taylor series near $\xi=0$, and noting that $t(\xi)$ is an odd function, making the complex exponential a mix of even and odd functions, the first relevant correction term in our regime of interest will be given by: 
\begin{equation}
\tilde{h}_{ij}^{(2)}(\omega) \approx \frac{4G}{r} \frac{1}{2}F''_{ij}(0) \int_{-\infty}^{\infty} d\xi\, ~\xi^2\exp\left(i \frac{\nu}{6} \xi^3\right).
\label{eq:integral_cubic_2}
\end{equation}
Rescaling and focusing only on the integral:
\begin{equation}
\begin{aligned}
I^{(2)}(\omega) &= \frac{1}{2}F_{ij}''(0) \int_{-\infty}^{\infty} \left(\frac{6}{\nu}\right)^{2/3} y^2 \cdot \left(\frac{6}{\nu}\right)^{1/3} dy\ e^{iy^3} \\
&= \frac{1}{2}F_{ij}''(0) \left(\frac{6}{\nu}\right) \int_{-\infty}^{\infty} dy\ y^2 e^{iy^3}.
\end{aligned}
\end{equation}
Using the known result for Fresnel-type integrals \cite{Gradshteyn:1943cpj}, we can show that the integral on $y$ vanishes in the sense of principal value.
So indeed the subdominant contribution at the $\xi^3$ (parabolic) dominated regime in this case also cannot give rise to a $\frac{1}{\nu}$ pole.
\section{Explicit form of $I(\nu, \theta', \phi')$} \label{sec:IFunc}
The function $I(\nu, \theta', \phi')$, as mentioned in Eq. \eqref{eq:interaction_term}, in terms of $\nu, \theta'$ and $\phi'$ is:
\begin{widetext}
\begin{equation}
    \begin{split}
        \mathrm{I}(\nu, \theta', \phi') &= \frac{1}{16} [8(2 (2 q_{xy}^2 - q_{xx} q_{yy} + (-2 q_{xy}^2 + q_{xx} q_{yy} + (2 q_{xy}^2 + q_{xx} q_{yy}) \cos^2 \theta') \cos^2 \phi') \sin^2 i \sin^2 \theta' \\ &+ q_{yy}^2 \sin^4 i \sin^4 \theta' + q_{xx}^2(-1 + \cos^2 \phi' \sin^2 \theta')^2) + 32 \cos i \cos \theta' \sin i \sin \theta'(-2 q_{xy}^2 \\ &+ ((2 q_{xy}^2 + q_{xx} q_{yy}) \cos^2 \phi'- q_{yy}^2 \sin^2 i) \sin^2 \theta') \sin \phi' + 8 q_{xx} q_{yy} \sin 2i \sin 2\theta' \sin \phi' \\ &+ 3 q_{yy}^2 \sin^2 2i \sin^2 2\theta' \sin^2 \phi' + 32 q_{yy}^2 \cos^3 i \cos \theta' \sin i \sin \theta' \sin \phi'(-1 + \sin^2 \theta' \sin^2 \phi') 
        \\ &+ 8 q_{yy}^3 \cos^4 i(-1 + \sin^2 \theta' \sin^2 \phi')^2 + 16 \cos^2 i(-((2 q_{xy}^2 - q_{xx} q_{yy})(-1 + \sin^2 \theta' \sin^2 \phi')) \\ &+ \cos^2 \phi' \sin^2 \theta'(-2 q_{xy}^2 + q_{xx}q_{yy} + q_{yy}^2 \sin^2 i + (2 q_{xy}^2 + q_{xx} q_{yy}) \sin^2 \theta' \sin^2 \phi'))].
    \end{split}
\end{equation}    
\end{widetext}
This $I(\nu, \theta', \phi')$ consists of polynomials of the Airy function and its derivative and also depends on the orientation angle of the binary $i$. Note that the fractional pole order in $\mathrm{I}(\nu, \theta', \phi')$ is of $\nu^{-2/3}$ which comes from the $q_{xx}^2$ term.

%

\end{document}